\documentclass[12pt]{article}
\usepackage[utf8]{inputenc}
\usepackage{slashed,amsmath,amsfonts,amssymb,bbm}
\usepackage[a4paper, total={6in, 8in}]{geometry}
\usepackage{comment}
\usepackage{graphicx}
\usepackage{caption}
\usepackage{subcaption}
\usepackage{hyperref}
\usepackage{cite}

\usepackage{array,multirow,float}
\usepackage{color}
\usepackage{mmacells}
\usepackage{colortbl}
\usepackage{subcaption}
\usepackage{threeparttable}
\hypersetup{
	colorlinks=true,
	linkcolor=magenta,
	filecolor=blue,      
	urlcolor=magenta
}
\usepackage[most]{tcolorbox}
\usepackage{tabularx}
\usepackage{booktabs}
\usetikzlibrary{patterns,shadows}
\tcbset{
	bgtable/.style={
		enhanced jigsaw,opacityback=.9,
		watermark graphics=#1
		}
}
\usepackage{indentfirst}

\newcommand{\ii}{\ensuremath{\mathrm{i}}}

\newcommand{\Gtil}{\ensuremath{\Tilde{G}}}
\newcommand{\Wtil}{\ensuremath{\widetilde{W}}}
\newcommand{\Btil}{\ensuremath{\widetilde{B}}}

\newcommand{\Ztil}{\ensuremath{\widetilde{Z}}}

 \newlength{\wdth}

\makeatletter

\begin{document}
\vspace{1cm}
{\flushleft{IFT-UAM/CSIC-23-64}}
\begin{figure}[h]
	\vspace{-2.9cm}
	\hspace*{\fill}
	\subcaptionbox*{}
	[.4\linewidth]%
	{\includegraphics[scale=0.12]{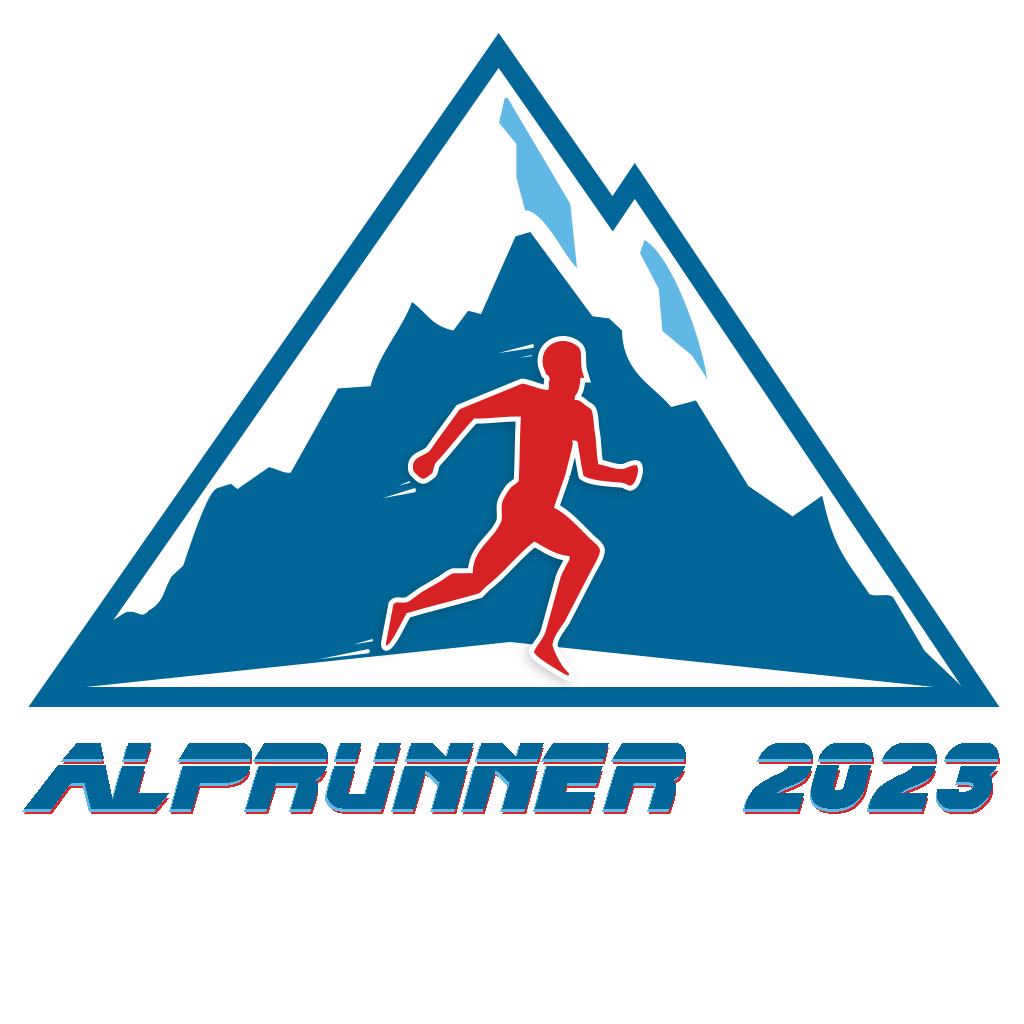}}
\end{figure}
\vspace{-1.3cm}
\begin{center}

{\Large\sc Running beyond ALPs:\\ shift-breaking and CP-violating effects}

\vspace{1cm}
\textbf{
Supratim Das Bakshi$^{\,a}$\footnote[1]{\url{sdb@ugr.es}}, Jonathan Machado-Rodr\'iguez$^{\,b}$\footnote[2]{\url{jonathan.machado@uam.es}}, Maria
Ramos$^{\,b}$\footnote[3]{\url{maria.pestanadaluz@uam.es}}}\\ 
\vspace{0.6cm} 
$^{\,a}$ \emph{CAFPE and Departamento de F\'isica Te\'orica y del Cosmos,\\ Universidad de Granada,
Campus de Fuentenueva, E–18071 Granada, Spain}

$^{\,b}$\emph{Departamento de F\'isica Te\'orica and Instituto de F\'isica Te\'orica IFT-UAM/CSIC,\\Universidad Aut\'onoma de Madrid, Cantoblanco, E-28049, Madrid, Spain}
\end{center}
\vspace{0.1cm}
\begin{abstract}
 We compute the renormalization group equations (RGEs) of the Standard Model effective field theory (EFT) extended with a real scalar singlet, up to dimension-five and one-loop accuracy. We compare our renormalization results with those found in the shift-symmetry preserving limit, which characterizes axion-like particles (ALPs). 
The matching and running equations below the electroweak scale are also obtained, 
including the mixing effects in the scalar sector.  
Such mixing leads to interesting phenomenological consequences that are absent in the EFT at the renormalizable level, 
namely new correlations among the triplet and quartic Higgs couplings are predicted.  
All RGEs obtained in this work are implemented in a new Mathematica package -- \texttt{ALPRunner}, together with functions to solve the running numerically for an arbitrary set of UV parameters. As an application, we 
obtain electric dipole moment constraints on particular regions of the singlet parameter space, and quantify the level of shift-breaking in these regions.
 %

%
\end{abstract}

\newpage

\tableofcontents

\vspace{1cm}

\section{Introduction}

\indent	  

Singlet scalars are one of the most promising candidates of beyond the Standard Model (BSM) physics. 
Not only can they provide solutions to long-standing puzzles of the SM, but their elusive nature can also explain their challenging detection, in spite of the increasing experimental efforts which span a wide range of energies and sophisticated techniques~\cite{Bauer:2017ris}.

If a new scalar particle is ever discovered, 
compelling questions will be addressed which could impact significantly our knowledge of Nature. 
One of these questions is how much the exotic scalar can interact with the only other we know, the Higgs boson. If the Higgs is afterall a composite particle, a plausible solution to the electroweak (EW) hierarchy problem~\cite{Kaplan:1983fs,Kaplan:1983sm}, it should be accompanied by several other scalar ``siblings'', most predicted to transform as singlets under the SM gauge group~\cite{Gripaios:2009pe,Chala:2012af,Vecchi:2013bja,Bellazzini:2014yua,Cacciapaglia:2019bqz}. The size of the scalar interactions, being UV dependent, could therefore provide valuable information about the underlying dynamics responsible for EW symmetry breaking (EWSB)~\cite{Gripaios:2009pe}.
Moreover, the Higgs-singlet portal is one of the simplest avenues to generate the dark matter relic density we observe~\cite{McDonald:1993ex,Burgess:2000yq,Arcadi:2019lka}, or even a first order phase transition, 
in turn connected with the possibility of explaining baryogenesis~\cite{Profumo:2007wc,Espinosa:2011eu,Chala:2016ykx,Ellis:2022lft}. 
The amount of CP violation the singlet carries would be then a key property to determine the viability of this scenario.
Another prominent question to face will be 
how much the new singlet looks like an 
axion-like particle (ALP). The existence of ALPs is a common prediction of more fundamental constructions, like string theory~\cite{WITTEN1984351,BANKS1996173}. They are furthermore motivated by another fine-tuning problem in the SM, the strong CP problem, which is the fine adjustment one is required to make (of more than ten orders of magnitude) to explain the absence of CP-violation in the QCD vacuum. To successfully solve this problem, the singlet interactions would have to preserve a shift-symmetry of very high quality~\cite{Peccei:1977ur,Peccei:1977hh,Weinberg:1977ma,Wilczek:1977pj}.

At the renormalizable level, the answers to these questions lie on measuring the strength of only a few singlet interactions with the SM particles. 
For instance, in the absence of mixing, the communication between a scalar singlet and the SM is entirely dictated by triplet and quartic interactions which are in turn related by the Higgs vacuum expectation value (VEV). Instead, if the two scalars mix, the singlet 
can also develop universal couplings to the SM fermions which are Yukawa suppressed, and moreover incompatible with a shift-symmetry.

Effective corrections can however change the answer to these questions, while renormalization group (RG) running can affect their stability. Therefore, in this work, we present the one-loop renormalization group equations (RGEs) characterizing the evolution across energy scales of all singlet interactions with the SM, up to mass dimension five. 
%
%
The RGEs of the singlet effective field theory (EFT) were previously obtained in two particular limits:  ignoring CP-odd couplings (under the assumption that the singlet is a pseudoscalar)~\cite{Chala:2020wvs}; or
including only shift-preserving interactions, that is assuming the singlet is an ALP~\cite{Bauer:2020jbp,Bonilla:2021pvu}. Our work is therefore the most complete study of the singlet EFT presented in the literature so far, not only at the level of RGEs, but also 
in what concerns all effective contributions to the scalar mass mixing. 
We obtain the running of the Wilson coefficients both in the high-energy (HE) and in the low-energy (LE) EFTs, the latter resulting from integrating out the heavy top quark, the Higgs and the $Z$ and $W$ gauge
bosons. The matching between the two theories is also obtained at tree level, including the effects which arise from scalar mixing. The shift-symmetric limit of our results is commented throughout the work.

As an application, we obtain the electric dipole moment (EDM) constraints on particular UV scenarios taking the full running of couplings into account. In particular, we set constraints on the CP violation arising from a CP-even BSM scenario, due to the mixing with the Yukawa couplings of the SM. We also exemplify how the experimental constraints of a generic singlet can be directly compared with those of an ALP. This comparison is based on the 
analysis presented in Ref.~\cite{Bonnefoy:2022rik}, where the shift-symmetry flavour invariants induced by the singlet-fermionic couplings were constructed.

All the results obtained in this work are provided in a new Mathematica package, \texttt{ALPRunner}~\cite{ALPRunner}, that can be used to study in detail the phenomenology of the singlet EFT. The main functionalities are described in App.~\ref{app:toolbox}. Using this tool, it is straightforward for the user to find the whole set of EFT parameters generated at an IR scale given the inputs defined in the UV. A procedure is also implemented such that, given low-energy bounds, constraints can be placed on the UV parameters of the theory.

The structure of this work can be easily understood from the table of contents.

\section{The complete singlet EFT}\label{sec:theory}

The renormalisable Lagrangian of the SM extended with a real~{scalar} singlet\,$(s)$ reads~\cite{OConnell:2006rsp,Barger:2007im,Gripaios:2016xuo}:
\begin{align}
	\mathcal{L}_{\text{SM}+s} &= \frac{1}{2} (\partial_{\mu} s) (\partial^{\mu} s) -\frac{1}{2} m_s^2 s^2 -\frac{\kappa_{s}}{3!}s^3 - \frac{\lambda_{s}}{4!} s^4 - \kappa_{s\phi} s \phi^\dagger \phi - \frac{\lambda_{s \phi}}{2} s^2 \phi^\dagger \phi  \nonumber\\
	& - \frac{1}{4} G^A_{\mu\nu} G^{A \mu\nu} - \frac{1}{4} W^a_{\mu\nu} W^{a \mu\nu} - \frac{1}{4} B_{\mu\nu} B^{\mu\nu} + \theta_{\rm QCD} \frac{g_s^2}{32 \pi^2} G^A_{\mu\nu} \widetilde{G}^{A \mu\nu}\nonumber \\
& + \sum\limits_{\psi=q,l,u,d,e} \overline{\psi}^\alpha i \slashed{D} \psi^\alpha -\left(y_{\alpha\beta}^u \overline{q_L}^\alpha \tilde{\phi} u_R^\beta + y_{\alpha\beta}^d \overline{q_L}^\alpha \phi d_R^\beta + y_{\alpha\beta}^e \overline{l_L}^\alpha \phi e_R^\beta +\text{h.c.} \right)\nonumber \nonumber \\
	&+(D_{\mu} \phi)^\dagger (D^{\mu} \phi) + \mu^2 \phi^\dagger \phi - \lambda (\phi^\dagger \phi)^2  \,,
\label{eq:renorm}
\end{align}
where we have used $e_R,\,u_R$ and $d_R$ to denote the right-handed (RH) leptons and quarks; while $l_L$ and $q_L$ are the left-handed (LH) counterparts. In turn, $ y^u $, $ y^d $, and $ y^e $ are the SM Yukawa matrices. We represent the Higgs doublet by $\phi = (\phi^+,\,\phi^0)^T$ and its charge conjugate by $\widetilde{\phi} = i \,  \sigma_2 \phi^\ast$. $ G^A_{\mu\nu} $, $ W^a_{\mu\nu} $, and $ B_{\mu\nu} $ are the field strength tensors of the SM gauge groups $ SU(3)_C $ , $ SU(2)_L $, and $ U(1)_Y $, respectively. The corresponding gauge couplings are denoted by $g_i$, with $i=1,2,3$. The dual strength tensor of the gluon field is defined as $\widetilde{G}^{A,\mu\nu} \equiv \frac{1}{2} \epsilon_{\mu\nu\rho \sigma} G^{A,\rho\sigma}$ and similarly for the other gauge fields.
We use the minus sign convention for the covariant derivative.

At dimension-five and assuming lepton number conservation, a minimal basis for the SM+$s$ EFT is given by the following operators~\cite{Chala:2020wvs}:
\begin{align}
\mathcal{L}_{{\rm SM}+s}^{\rm eff} =&~ \mathcal{L}_{\text{SM}+s}+ s \bigg[ i \,  \overline{q_L} a_{su\phi} \widetilde{\phi} u_R + i \,  \overline{q_L} a_{sd\phi} {\phi} d_R +i \,  \overline{l_L} a_{se\phi} {\phi} e_R + \text{h.c.} \bigg]  + a_{s^5} s^5 \nonumber \\ 
& + a_{s ^3} s^3 (\phi^\dagger \phi) + a_{s} s (\phi^\dagger \phi)^2  
+  a_{s\widetilde{G}} s G_{\mu\nu}^{A}\widetilde{G}^{A\mu\nu} + a_{s\widetilde{W}} s W_{\mu\nu}^{a}\widetilde{W}^{a\mu\nu}  \nonumber \\
& +  a_{s\widetilde{B}} s B_{\mu\nu}\widetilde{B}^{\mu\nu}+ a_{s{G}} s G_{\mu\nu}^{A}{G}^{A\mu\nu} + a_{s{W}} s W_{\mu\nu}^{a}{W}^{a\mu\nu} +  a_{s{B}} s B_{\mu\nu}{B}^{\mu\nu}\,,
\label{eq:basis}
\end{align}
where $a_{s\psi\phi}$ are complex matrices in flavour space to account for both CP-conserving and CP-violating interactions. In our notation, all $a_i \equiv a_i^0/\Lambda$ are dimension-full couplings, with $\Lambda$ denoting the cutoff scale up until which the EFT is a valid expansion. In what follows, we assume $s$ is a pseudoscalar. 
The previous EFT and the results to obtain next are however valid beyond this assumption. 

To order $v/\Lambda$, the dimension-five Wilson coefficients are only renormalized by other dimension-five couplings:
\begin{equation}
\label{eq:AD-5}
16 \pi^2 \mu \frac{\text{d} a_i}{\text{d} \mu} = \gamma_{ij}^{(1)} a_j\,.
\end{equation}
The anomalous dimensions $\gamma^{(1)}$ were computed in Ref.~\cite{Chala:2020wvs} for the CP-even sector in isolation. We generalize Eq.~\ref{eq:AD-5} by including the contributions from CP-odd couplings. In this more general scenario, dimension-five terms renormalize lower dimensional interactions as well, due to the presence of light mass scales in the theory.
This effect is absent assuming only CP-even interactions as, in this case, there is a $\mathcal{Z}_2$ symmetry acting on $s\to -s$, under which the (non-) renormalizable sector of the EFT is (odd) even. 

Considering renormalizable couplings, we therefore expect, schematically, the following structure: 
\begin{align}
\label{eq:AD-3}
16 \pi^2 \mu \frac{\text{d} \kappa_i}{\text{d} \mu} & = \gamma_{ij}^{(2)} \kappa_j + \gamma_{ij}^{(3)} m^2 a_j \, , \\
16 \pi^2 \mu \frac{\text{d} \lambda_i}{\text{d} \mu} & = \gamma_{ij}^{(4)} \lambda_j + \gamma_{ijk}^{(5)} \kappa_j a_k \, .
\end{align}
where $m=m_s,\,\mu$. The RGEs of the CP-odd couplings in Eq.~\ref{eq:AD-3} are presented for the first time in this work, as well as the contributions stemming from $\gamma^{(5)}$.
\vspace{-0.1cm}

\subsection{A comparison with the ALP basis}~\label{sec:ALPbasis}
\vspace{-0.5cm}

In order to account for the most generic interactions, we did not impose shift-symmetry in the EFT defined in Eq.~\ref{eq:basis}, as commonly done in studies of axions and ALPs, which arise as the Goldstone bosons of a spontaneously broken global symmetry. In particular, the axion is a prediction of the Peccei-Quinn mechanism~\cite{Peccei:1977ur,Peccei:1977hh,Weinberg:1977ma,Wilczek:1977pj} to  
solve the strong CP problem, which requires an exact shift symmetry in the axion Lagrangian, broken at the quantum level by the anomalous interaction with gluons. 
Upon QCD confinement, such interaction generates a periodic potential for the axion, $V(s) = V(s+2\pi f_s)$, 
with $f_s$ denoting the field decay constant. Such potential also predicts a strict relation between the mass and decay constant for the axion field, that ALPs are not required to abide. 

Up to a bare mass term, the minimal set of ALP interactions can be described by the following EFT~\cite{GEORGI198673}:
\begin{align}
\label{eq:ALPL}
\mathcal{L}_{\rm ALP} & =\frac{1}{2} (\partial_\mu s)^2 +  \sum_{\psi}\frac{\partial_\mu s}{f_s}
\overline{\psi} c_\psi \gamma^\mu \psi + \sum_{X} c_X \frac{g_X^2}{16 \pi^2}  \frac{s}{f_s} X_{\mu\nu}\widetilde{X}^{\mu\nu} \,,
\end{align}
where $\psi$ runs over $q_L, l_L, u_R, d_R, e_R$, $c_\psi$ are hermitian
matrices in flavour space and $X=G,W,B$. %
When the gauge boson sector of this theory is considered in isolation, the $2\pi$ periodicity imposes that $c_X \in \mathbb{Z}$~\cite{Agrawal:2022yvu}. While the shift $s\to s+ 2\pi f_s$ modifies the interaction Lagrangian, the action is left invariant whenever this quantization condition is fulfilled.
It is well known that a chiral rotation produces, however, additional interactions which modify this condition~\cite{Fraser:2019ojt}. In other words,  the requirement $c_X  \in \mathbb{Z}$ follows from the ALP periodicity only in the basis of Eq.~\ref{eq:ALPL}.

Consequently, in this basis, the running of $c_X$ couplings is expected to vanish\,\footnote{This argument is valid up to the inclusion of a potential, whose truncation can break explicitly the ALP periodicity.}, as found in previous works~\cite{Chala:2020wvs,Bauer:2020jbp,Bonilla:2021pvu} (up to two-loop accuracy in the gauge couplings). On the other hand, there is no constraint on the running of the derivative couplings. 

In the interaction basis considered in this work (Eq.~\ref{eq:basis}), the singlet periodicity is explicitly broken. Therefore, no quantization constraint applies\,\footnote{There can be other reasons preventing the one-loop running of the couplings of a general scalar singlet.}.
The previous discussion still connects to our work in two main ways:
\begin{itemize}
\item The more restrictive running in the SM+$s$ EFT invariant under discrete ALP shifts constitutes an important case limit of our study, which should be recovered from the complete RGEs to be derived next.

\item Using the freedom to write the derivative interactions between the ALP and fermions in terms of chirality-flipping operators, it is straightforward to find the conditions to preserve shift-symmetry in the basis of Eq.~\ref{eq:basis}. Hence, the ALP phenomenology can be studied using the EFT basis of this work, even though it is not possible in general to map the couplings back to the chirality-preserving basis in Eq.~\ref{eq:ALPL}.
\end{itemize}
Indeed, by integrating by parts the second term in Eq.~\ref{eq:ALPL} and using the SM equations of motion (EOMs), we obtain the fermionic operators in Eq.~\ref{eq:basis} with the following coefficients~\cite{Chala:2020wvs}:
\begin{equation}
\label{eq:aSS}
a_{su\phi} = \frac{i}{f_s} \,  \left(y^{u}c_{u} - c_{q} y^u \right)\,,\quad a_{sd\phi}= \frac{i}{f_s} \,  \left(y^d c_{d} - c_{q} y^d \right)\,,\quad a_{se\phi} = \frac{i}{f_s} \,  \left({y^{e}} c_{e} - c_{l} y^{e} \right)\,.
\end{equation}
These three conditions are sufficient to preserve a continuous shift-symmetry at the perturbative level in our EFT, 
up to terms in the scalar potential. However these conditions are implicit and do not, in general, allow to identify the $c$-matrices given a set of $a_{s\psi\phi}$ couplings. 

Only under specific assumptions, such as Minimal Flavour Violation or in the presence of just one fermion generation coupled to the ALP, it is possible to trade completely $c_l,c_q$ by their RH counterparts and find one-to-one correspondence between the chirality-preserving and chirality-flipping couplings~\cite{Chala:2020wvs,Bauer:2020jbp,Bonilla:2021pvu}.
In those cases, to match the results derived using the basis in Eq.~\ref{eq:ALPL} and ours, shifts on the anomalous couplings also need to be taken into account, that arise from the axial anomaly equation~\cite{Brivio:2017ije}:
\begin{align}
\label{eq:cSS0}
      &a_{s\tilde{G}}  = \frac{g_3^2}{16 \pi^2 f_s} \bigg[c_G - \frac{1}{2} \text{Tr}\left(c_u + c_d - 2 c_q\right) \bigg]\,, \\
            &a_{s\tilde{W}}  = \frac{g_2^2}{16 \pi^2 f_s} \bigg[c_W + \frac{1}{2} \text{Tr}\left(3  c_q + c_l \right) \bigg]\,, \\
            \label{eq:cSS}
                        &a_{s\tilde{B}}  = \frac{g_1^2}{16 \pi^2 f_s} \bigg[c_B - \text{Tr}\left( \frac{4}{3} c_u + \frac{1}{3} c_d - \frac{1}{6}  c_q + c_e -\frac{1}{2} c_l \right) \bigg] \,.
\end{align}
Assuming this set of boundary conditions (Eqs.~\ref{eq:aSS}--\ref{eq:cSS}), the results obtained with any of the two bases must agree within an accuracy of $\mathcal{O}(\alpha/\Lambda)$.

Finally, we remark that the general basis adopted in this work can be used to study the fate of shift-symmetric scenarios under running, without having to refer implicitly to the matrices $c_\psi$. This has become possible with the work presented in Ref.~\cite{Bonnefoy:2022rik}, where the complete set of flavour invariants associated to the singlet shift-symmetry was constructed, and shown to form a closed set under RG evolution.

\section{High-energy anomalous dimensions}\label{sec:adm}

We renormalize the SM+$s$ EFT by computing the divergences in the basis of independent Green's functions of Ref.~\cite{Chala:2020wvs}; see App.~\ref{sec:greensbases}. 
We use the background field method (BFM) and work in the Feynman gauge with $\text{d} = 4-2\epsilon$ spacetime dimensions. The computations are based on dedicated routines that rely on \texttt{FeynRules}~\cite{Alloul:2013bka}, \texttt{FeynArts}~\cite{Hahn:2000kx}, and \texttt{FormCalc}~\cite{Hahn:2016ebn}. All of our results have been cross-checked with \texttt{matchmakereft}~\cite{Carmona:2021xtq}, from which we have obtained the $\mu^2$ contributions.

\vspace{0.1cm}

The one-loop divergences of the renormalizable couplings read:
\begin{align}
    m^{\prime\, 2}_s & =-\frac{1}{32\pi^2 \epsilon}\Bigg(\lambda_s m_s^2 + \kappa_s^2 + 4 \kappa_{s\phi}^2 -{4 \lambda_{s\phi} \mu^2} \Bigg) \,,\\
    \label{eq.divergence-muphi}
    \mu^{\prime\, 2} & = \frac{1}{32\pi^2 \epsilon}\Big( 2 \kappa_{s\phi}^2 + \lambda_{s\phi} m_s^2 + {\frac{1}{2} \left(g_1^2 + 3 g_2^2 -12 \lambda \right) \mu^2	}	\Big)  \,,\\
    \label{eq.divergence-kappas}
    \kappa'_{s} & = -\frac{3}{4\pi^2 \epsilon}\left( \frac{1}{8}\kappa_{s} \lambda_{s} + \frac{1}{2}\kappa_{s\phi}\lambda_{s\phi} -5 a_{s^5}m_{s}^2 {+ a_{s^3} \mu^2} \right)  \,,\\
     \label{eq.divergence-kappasphi}
    \kappa'_{s\phi} &= -\frac{1}{32\pi^2 \epsilon}\left( \kappa_{s} \lambda_{s\phi} - \frac{1}{2}\kappa_{s\phi}\left[g_1^2 +3g_2^2 - 8({ 3} \lambda + \lambda_{s\phi}) \right] - 6a_{s^3}m_{s}^2	{+12 a_s \mu^2} \right) \,,\\
        \label{eq.divergence-lambdas}
    \lambda'_{s} & = -\frac{1}{4\pi^2 \epsilon}\left( \frac{3}{8} \lambda_{s}^2 + \frac{3}{2}\lambda_{s\phi}^2 -12\left( \kappa_{s\phi} a_{s^3} + 5 \kappa_{s} a_{s^5} \right) \right)  \,,\\
      \label{eq.divergence-lambdasphi}
    \lambda'_{s\phi} & = -\frac{1}{8\pi^2 \epsilon}\left\{ \lambda_{s \phi} \left[ \lambda_{s\phi} + \frac{1}{4}\lambda_{s} +  3  \lambda -\frac{1}{8}\left( g_1^2 + 3g_2^2 \right) \right] \right. \nonumber \\ &~~~~ \left. - \bigg[ 3\kappa_{s} a_{s^3} +{6\left( a_{s} + a_{s^3} \right)}\kappa_{s\phi}  \bigg]  \right\}  \,,\\
     \label{eq.divergence-lambdaphi}
    \lambda' & = -\frac{1}{32\pi^2 \epsilon}\Bigg\{\frac{3}{8}\left(g_1^4 + 3 g_2^4 + 2g_1^2 g_2^2 \right) - \left(g_1^2 + 3g_2^2\right)\lambda + {24} \lambda^2 + \frac{\lambda_{s\phi}^2}{2} \nonumber \\
   & ~~~~ -8 \kappa_{s\phi} a_s 	- 2\text{Tr}\Big[ 3\left( y_u^\dagger y_u y_u^\dagger y_u + y_d^\dagger y_d y_d^\dagger y_d \right) + y_e^\dagger y_e y_e^\dagger y_e \Big] \Bigg\}  \,,\\
    y'_{u} & = \frac{1}{16\pi^2 \epsilon}\left[ y_d y_d^\dagger y_u + \frac{1}{36}\left( 25 g_1^2 + 27g_2^2 +192 g_3^2\right)y_u + i \kappa_{s \phi} a_{s u \phi}\right]  \,,\\
    \label{eq.divergence-yd}
    y'_d & = \frac{1}{16\pi^2 \epsilon}\left[ y_u y_u^\dagger y_d + \frac{1}{36}\left( g_1^2 + 27g_2^2 +192 g_3^2\right)y_d + i \kappa_{s \phi} a_{s d \phi}\right]  \,,\\
    \label{eq.divergence-ye}
    y'_e & = \frac{3}{64\pi^2 \epsilon}\left[\left( 3g_1^2 +g_2^2 \right)y_e  + \frac{4i}{3} \kappa_{s \phi} a_{s e \phi} \right]  \,.
\end{align}
\vspace{0.05cm}

\noindent As we work in BFM approach, 
the divergences of the gauge couplings are automatically fixed by the wave function renormalization (WFR) factors of the gauge bosons~\cite{Abbott:1981ke}. The latter are not modified by the presence of CP-odd terms and can therefore be read from Ref.~\cite{Chala:2020wvs}. 

We obtain the following divergences for the effective couplings\,\footnote{In several of the following terms, the unit matrix in flavour space is left implicit.}:
\begin{align}
\label{eq.divergence-s5}
a'_{s^5} & = -\frac{1}{16\pi^2 \epsilon}\left( 5 \lambda_{s} a_{s^5} + \lambda_{s\phi} a_{s^3} \right)  \,,\\
\label{eq.divergence-s3phi}
    a'_{s^3} & = -\frac{1}{32\pi^2 \epsilon}\left[ \left( 12 \lambda + 3 \lambda_{s} + 12 \lambda_{s\phi} -\frac{g_1^2}{2} -\frac{3g_2^2}{2}\right)  a_{s^3} + 6 \lambda_{s\phi} a_{s} + 20 \lambda_{s\phi}a_{s^5}\right]  \,,\\
    \label{eq.divergence-sphi2}
    a'_{s} & = -\frac{1}{32\pi^2 \epsilon}\Bigg\{ \left( 48 \lambda  + 8 \lambda_{s\phi} -g_1^2 -3g_2^2 \right) a_{s} -3 \left( g_1^4 +g_1^2 g_2^2 \right)a_{sB} -3 \left( 3 g_2^4 + g_1^2 g_2^2 \right)a_{sW} \nonumber  \\ &~~~+ 6\lambda_{s\phi}a_{s^3} +  8\text{Im}\left\{ \text{Tr}\left[3\left(y_d^\dagger y_d y_d^\dagger a_{sd\phi} + y_u^\dagger y_u y_u^\dagger a_{su\phi} \right) + y_e^\dagger y_e y_e^\dagger a_{se\phi}\right] \right\} \Bigg\}  \,,\\
         \label{eq.divergence-suphi}
    a'_{su\phi} & = -\frac{1}{16\pi^2 \epsilon}\left\{ a_{su\phi}\left[ \lambda_{s\phi} - \left( \frac{25 g_1^2}{36} + \frac{3g_2^2}{4} + \frac{16g_3^2}{3} \right) \right] - y_d y_d^\dagger a_{su\phi} -a_{sd\phi}y_d^\dagger y_u \right. \nonumber \\ &~~~+\left. y_d a_{sd\phi}^\dagger y_u - i\left(\frac{4 g_1^2}{3} a_{sB} + 16 g_3^2 a_{sG} \right) y_{u} \right\} \,, \\
     \label{eq.divergence-sdphi}
    a'_{sd\phi} &= -\frac{1}{16\pi^2 \epsilon}\left\{ a_{sd\phi}\left[ \lambda_{s\phi} - \left( \frac{g_1^2}{36} + \frac{3g_2^2}{4} + \frac{16g_3^2}{3} \right) \right] - y_u y_u^\dagger a_{sd\phi} -a_{su\phi}y_u^\dagger y_d +\right. \nonumber\\ &~~~+\left. y_u a_{su\phi}^\dagger y_d + i\left(\frac{2}{3} g_1^2 a_{sB} - 16 g_3^2 a_{sG} \right) y_d\right\} \,,\\
    \label{eq.divergence-sephi}
    a'_{se\phi} & = -\frac{1}{16\pi^2 \epsilon}\left\{ a_{se\phi}\left[ \lambda_{s\phi} - \left( \frac{9 g_1^2}{4} + \frac{3g_2^2}{4} \right) \right] - 6 i g_1^2 a_{sB} y_e \right\}  \,,\\
    \label{eq.divergence-sq}
    r'_{sq} & = \frac{1}{32\pi^2 \epsilon}\left[ a_{sd\phi} y_d^\dagger + a_{su\phi} y_u^\dagger - \left( \frac{g_1^2}{3} a_{s\widetilde{B}} + 9 g_2^2 a_{s\widetilde{W}} +16 g_3^2 a_{s\widetilde{G}} \right) \right] \,,\\
    \label{eq.divergence-sl}
    r'_{sl} & = \frac{1}{32\pi^2 \epsilon}\left( a_{se\phi} y_e^\dagger -3 g_1^2 a_{s\widetilde{B}} -9 g_2^2 a_{s\widetilde{W}}  \right) \,,\\
    \label{eq.divergence-su}
    r'_{su} & = -\frac{1}{16\pi^2 \epsilon} \left(a_{su\phi}^\dagger y_u - \frac{8 g_1^2}{3} a_{s\widetilde{B}} -8 g_3^2 a_{s\widetilde{G}} \right) \,,\\
    \label{eq.divergence-sd}
    r'_{sd} & = -\frac{1}{16\pi^2 \epsilon}\left(a_{sd\phi}^\dagger y_d - \frac{2}{3} g_1^2 a_{s\widetilde{B}} -8 g_3^2 a_{s\widetilde{G}} \right) \,,\\
    \label{eq.divergence-se}
    r'_{se} & = -\frac{1}{16\pi^2 \epsilon}\left(a_{se\phi}^\dagger y_e -6 g_1^2 a_{s\widetilde{B}}\right)  \,,\\    
         \label{eq.divergence-sphibox}
    r'_{s\phi\Box} & = -\frac{1}{16\pi^2 \epsilon}\left\{ \text{Tr}\left[ 3y_d a_{sd\phi}^\dagger -3 y_u^\dagger a_{su\phi} + y_e a_{se\phi}^\dagger  \right] + \frac{3}{2} i \left[ g_1^2 a_{sB} + 3 g_2^2 a_{sW} \right] \right\} \,,
\end{align}
where the $r$-couplings are associated to redundant operators which will be redefined away; see App.~\ref{sec:greensbases}. All other counterterms vanish up to the EFT and loop order we are interested in. 

The divergences obtained above can be projected onto the minimal basis of Eq.~\ref{eq:basis} using suitable field redefinitions, which can be implemented via the EFT EOMs within $\mathcal{O}(1/\Lambda)$ accuracy. In this procedure, we neglect $\mathcal{O}(\alpha/4\pi)$ corrections that arise from the anomaly equations which relate chirality-preserving operators with the chirality-flipping ones in our basis; see Eqs.~\ref{eq:cSS0}-\ref{eq:cSS}. This would lead to $\mathcal{O}(\alpha/4\pi)^2$ corrections which are beyond the scope of our analysis. 
The on-shell results are obtained after the following replacements\,\footnote{These replacements differ from those in Ref.~\cite{Chala:2020wvs}, where only the CP-even components of the flavour couplings were considered.}:
    \begin{align}
    \label{eq.redundancy-kappas}
    \kappa'_{s} \rightarrow & \kappa'_{s} + 6 m_s^2 r'_{s\Box} \nonumber\\
    = &-\frac{3}{4\pi^2 \epsilon}\left[ \frac{1}{8}\kappa_{s} \lambda_{s} + \frac{1}{2}\kappa_{s\phi}\lambda_{s\phi} -5 a_{s^5}m_{s}^2 {+\mu^2 a_{s^3}} \right]  \,,\\
     \label{eq.redundancy-kappasphi}
    \kappa'_{s\phi} \rightarrow & \kappa'_{s\phi} -2 \text{Im}\left[ r'_{s\phi\Box}\right] \mu^2 + r'_{\phi s\Box}m_s^2 \nonumber \\
    = & -\frac{1}{32\pi^2 \epsilon}\Bigg\{ \kappa_{s} \lambda_{s\phi} - \frac{1}{2}\kappa_{s\phi}\left[g_1^2 +3g_2^2 - 8({3}\lambda + \lambda_{s\phi}) \right] - 6a_{s^3}m_{s}^2 {+ 12 \mu^2 a_s} \nonumber \\
    - & 6 \mu^2 \left( g_1^2 a_{sB} + 3 g_2^2 a_{sW}\right) +4 \mu^2 \text{Im}\Bigg[ \text{Tr}\left[ 3\left(y_u^\dagger a_{su\phi} - y_d a_{sd\phi}^\dagger \right) - y_e a_{se\phi}^\dagger \right]\Bigg] \Bigg\} \,,\\
        \label{eq.redundancy-lambdas}
    \lambda'_{s} \rightarrow & \lambda'_{s} + 12 \kappa_s r'_{s\Box}\nonumber \\
    =& -\frac{1}{4\pi^2 \epsilon}\left[ \frac{3}{8} \lambda_{s}^2 + \frac{3}{2}\lambda_{s\phi}^2 -12\left( \kappa_{s\phi} a_{s^3} + 5 \kappa_{s} a_{s^5} \right) \right]  \,,\\
      \label{eq.redundancy-lambdasphi}
    \lambda'_{s\phi}\rightarrow & \lambda'_{s\phi} + 2 \kappa_{s\phi} r'_{s\Box} - 4 \kappa_{s\phi} \text{Im}\left[ r'_{s\phi\Box} \right] + \kappa_s r'_{\phi s\Box} \nonumber \\
    = & -\frac{1}{8\pi^2 \epsilon}\Bigg\{ \lambda_{s\phi} \left[ \lambda_{s\phi} + \frac{1}{4}\lambda_{s} + {3} \lambda -\frac{1}{8}\left( g_1^2 {+} 3g_2^2 \right) \right] - \bigg[ 3\kappa_{s} a_{s^3} +2\left( {3} a_{s} + 3 a_{s^3} \right)\kappa_{s\phi} \bigg] \nonumber \\
    	- & 3 \kappa_{s\phi} \left( g_1^2 a_{sB} + 3 g_2^2 a_{sW}\right) +2 \kappa_{s\phi} \text{Im}\Bigg[ \text{Tr}\left[ 3\left(y_u^\dagger a_{su\phi} - y_d a_{sd\phi}^\dagger \right) - y_e a_{se\phi}^\dagger \right]\Bigg] \Bigg\}	\,,\\
 \label{eq.redundancy-lambdaphi} 
    \lambda' \rightarrow & \lambda' + \kappa_{s\phi} r'_{\phi s\Box} \nonumber \\
    = & -\frac{1}{32\pi^2 \epsilon}\Bigg\{\frac{3}{8}\left(g_1^4 + 3 g_2^4 + 2g_1^2 g_2^2 \right) - \left(g_1^2 + 3g_2^2\right)\lambda + {24} \lambda^2 + \frac{\lambda_{s\phi}^2}{2} {- 8 \kappa_{s\phi} a_s}\nonumber \\
   & ~~~~ - 2\text{Tr}\Big[ 3\left( y_u^\dagger y_u y_u^\dagger y_u + y_d^\dagger y_d y_d^\dagger y_d \right) + y_e^\dagger y_e y_e^\dagger y_e \Big] \Bigg\}  \,,
	\end{align}
	\begin{align}
    \label{eq.redundancy-suphi}
    a'_{su\phi} \rightarrow & a'_{su\phi} - r'_{s\phi\Box} y_u - r'_{sq} y_u + y_u (r'_{su})^\dagger \nonumber \\
    =& -\frac{1}{(4\pi)^2 \epsilon} \Bigg\{ \left[\lambda_{s\phi} -\frac{25}{36}g_1^2 - \frac{3}{4}g_2^2 -\frac{16}{3}g_3^2\right]a_{su\phi} + \text{Tr}\Big[3 \left(y_u^\dagger a_{su\phi} - y_d a_{sd\phi}^\dagger \right)- y_e a_{se\phi}^\dagger \Big]y_u \nonumber \\
    -& \frac{1}{2} a_{sd\phi}y_d^\dagger y_u + \frac{1}{2} a_{su\phi}y_u^\dagger y_u + y_d a_{sd\phi}^\dagger y_u - y_d y_d^\dagger a_{su\phi}+ y_u y_u^\dagger a_{su\phi} \nonumber
    \\-& \Big[\frac{17}{6}\left( i a_{sB}+a_{s\Btil} \right)g_1^2+ \frac{9}{2}\left( i a_{sW}+a_{s\Wtil}\right)g_2^2 + 16\left( ia_{sG}+a_{s\Gtil}\right) \Big]y_u\Bigg\} \,, 	\end{align}
	\begin{align}
    \label{eq.redundancy-sdphi}
    a'_{sd\phi} \rightarrow & a'_{sd\phi} + r'^{*}_{s\phi\Box} y_d - r'_{sq} y_d + y_d r_{sd}^\dagger \nonumber \\
    =& -\frac{1}{(4\pi)^2 \epsilon} \Bigg\{ \left[\lambda_{s\phi} -\frac{1}{36}g_1^2 - \frac{3}{4}g_2^2 -\frac{16}{3}g_3^2\right]a_{sd\phi} + \text{Tr}\Big[3\left(y_d^\dagger a_{sd\phi} - y_u a_{su\phi}^\dagger\right) + y_e^\dagger a_{se\phi} \Big]y_d  \nonumber 
    \\ -& \frac{1}{2} a_{su\phi}y_u^\dagger y_d + \frac{1}{2} a_{sd\phi}y_d^\dagger y_d + y_u a_{su\phi}^\dagger y_d - y_u y_u^\dagger a_{sd\phi}+ y_d y_d^\dagger a_{sd\phi} \nonumber 
    \\-& \Big[\frac{5}{6}\left( i a_{sB}+a_{s\Btil} \right)g_1^2+ \frac{9}{2}\left( i a_{sW}+a_{s\Wtil}\right)g_2^2 + 16\left( ia_{sG}+a_{s\Gtil}\right) \Big]y_d\Bigg\} \,, 	\end{align}

	\begin{align}
    \label{eq.redundancy-sephi}
    a'_{se\phi} \rightarrow & a'_{se\phi} + r'^{*}_{s\phi\Box} y_e - r'_{sl} y_e + y_e r_{se}^\dagger \nonumber \\
    =& -\frac{1}{(4\pi)^2 \epsilon} \Bigg\{ \left[\lambda_{s\phi} -\frac{9}{4}g_1^2 - \frac{3}{4}g_2^2 \right]a_{se\phi} + \text{Tr}\Big[3\left(y_d^\dagger a_{sd\phi} - y_u a_{su\phi}^\dagger \right)+ y_e^\dagger a_{se\phi} \Big]y_e \nonumber \\
    +& \frac{1}{2} a_{se\phi}y_e^\dagger y_e + y_e y_e^\dagger a_{se\phi} - \frac{3}{2} \Big[5\left( i a_{sB}+a_{s\Btil} \right)g_1^2+ 3\left( i a_{sW}+a_{s\Wtil}\right)g_2^2 \Big]y_e\Bigg\} \,, 
            \end{align}
            
    \begin{align}
    \label{eq.redundancy-s5}
    a'_{s^5} \rightarrow & a'_{s^5} - \frac{\lambda_s}{3!}r'_{s\Box} \nonumber \\
    =& {-\frac{1}{16\pi^2 \epsilon}}\left( 5 \lambda_s a_{s^5} + \lambda_{s\phi} a_{s^3} \right) \,,  \\
     \nonumber\\
        \label{eq.redundancy-s3}
    a'_{s^3} \rightarrow & a'_{s^3} + \left[\text{Im}\left( r'_{s\phi\Box}\right) - r'_{s\Box}\right]\lambda_{s\phi} - \frac{\lambda_s}{3!}r'_{\phi s \Box} \nonumber \\
    =& -\frac{1}{32\pi^2 \epsilon}\Bigg\{ \left( 12 \lambda  + 3 \lambda_{s} + 12 \lambda_{s\phi} -\frac{g_1^2}{2} -\frac{3g_2^2}{2}\right)  a_{s^3} +4 \lambda_{s\phi} \left({\frac{3}{2}}a_{s} + 5 a_{s^5} \right) \nonumber \\
    +& 3 \lambda_{s\phi} \left( g_1^2 a_{sB} + 3g_2^2 a_{sW}\right) + 2 \lambda_{s\phi} \text{Im}\Bigg[ \text{Tr}\Big[ 3\left(y_d a_{sd\phi}^\dagger - a_{su\phi}y_u^\dagger \right) + y_e a_{se\phi}^\dagger \Big] \Bigg] \Bigg\}  \,, 
    \end{align}
    \begin{align}
    \label{eq.redundancy-s}
    a'_{s} \rightarrow & a'_{s} + 4\lambda \text{Im}\left( r'_{s\phi\Box}\right) - \lambda_{s\phi} r'_{\phi s\Box} \nonumber \\
    =& -\frac{1}{32\pi^2 \epsilon}\Bigg\{ \left( 48 \lambda + 8 \lambda_{s\phi} -g_1^2 -3g_2^2 \right) a_{s} + 6\lambda_{s\phi}a_{s^3} -3 \left( g_1^4 +g_1^2 g_2^2 -4 \lambda g_1^2 \right)a_{sB} \nonumber \\ 
    -& 3 \left( 3 g_2^4 + g_1^2 g_2^2 -12 \lambda g_2^2 \right)a_{sW} +  8\text{Im}\Bigg[ \text{Tr}\Big[3\left(y_d^\dagger y_d y_d^\dagger a_{sd\phi} + y_u^\dagger y_u y_u^\dagger a_{su\phi} \right) \nonumber \\ 
    +& y_e^\dagger y_e y_e^\dagger a_{se\phi} + \lambda \left( 3 y_d a_{sd\phi}^\dagger - 3 y_u^\dagger a_{su\phi} + y_e a_{se\phi}^\dagger \right)\Big] \Bigg] \Bigg\}  \,.
    \end{align}

We can now obtain the anomalous dimensions of all couplings in the singlet EFT. Taking the CP-even limit of our results, we find exact agreement with those presented in Ref.~\cite{Chala:2020wvs}.
\begin{table}[t]
	\centering{}
	\renewcommand{\arraystretch}{1.9}
	\begin{tabular}{c|c c c c c c }
		\hline
		& $s^5$ & $s^3 \phi^\dagger \phi$ & $s (\phi^\dagger \phi)^2$ & $s\overline{\Psi_L} \phi \psi_R$ & $s XX$ & $s X \widetilde{X}$ \\ \hline
		$s^5$ & $\textcolor{blue}{\lambda_s}$ & $\lambda_{s\phi}$ & 0 & 0 & 0 & 0 \\
		$s^3 \phi^\dagger \phi$ & $\textcolor{blue}{\lambda_{s\phi}}$ & $\lambda_s + \textcolor{blue}{\lambda_{s\phi}} +\textcolor{blue}{\lambda} + y_t^2  $ & $\lambda_{s \phi}$ & $\lambda_{s\phi} y_t$ & $\lambda_{s\phi} g_2^2$ & 0 \\
		$s (\phi^\dagger \phi)^2 $ & 0 & $\lambda_{s\phi}$ & $\lambda_{s\phi}$ + $\textcolor{blue}{\lambda} +\textcolor{blue}{ y_t^2}$ & $\textcolor{blue}{ y_t^3 } + \textcolor{blue}{ \lambda y_t}  $ & $\textcolor{blue}{\lambda g_2^2}$ & 0 \\
		$s\overline{\Psi_L} \phi \psi_R$ & 0 & 0 & 0 & $\lambda_{s\phi}+\textcolor{blue}{y_t^2}$ & $ \textcolor{blue}{g_3^2 y_t}$ & $ \textcolor{blue}{g_3^2 y_t}$ \\
		$s XX$ & 0 & 0 & 0 & 0 & $\textcolor{blue}{g_3^2}$ & 0  \\
		$s X \widetilde{X}$ & 0 & 0 & 0 & 0 & 0 & $\textcolor{blue}{g_3^2}$ 
	\end{tabular}
	\caption{\it Structure of the dimension-five anomalous dimension matrix. Only BSM and the leading SM contributions are kept. For instance, $ y_t  $ (the top-quark Yukawa coupling) and $g_{2,3} $ are the largest contributions in their respective sub-classes. Terms in blue show the contributions that deviate significantly from naive dimensional analysis; see the text for details.}
	\label{tab:uv5}
\end{table}
\begin{table}[h]
	\centering{}
	\renewcommand{\arraystretch}{1.9}
	\begin{tabular}{c|c c c c c c}
		\hline
		& $s^5$ & $s^3(\phi^\dagger \phi)$ & $s (\phi^\dagger \phi)^2$ & $s\overline{\Psi_L} \phi \psi_R$ & $s XX$ & $s X \widetilde{X}$ \\ \hline
		$s^3$ & $\textcolor{blue}{m_s^2}$ & $\textcolor{blue}{\mu^2}$ &  $0$  & $0$ & $0$ & $0$\\
		$s(\phi^\dagger \phi)$ & $0$ & $m_s^2$ &  $\textcolor{blue}{\mu^2}$ & $y_t \mu^2$ & $ \textcolor{blue}{g_2^2 \mu^2} $ & $0$ \\
		$s^4$ & $\textcolor{blue}{\kappa_s}$ & $\textcolor{blue}{\kappa_{s\phi}}$ & $0$ & 0  & $0$ & $0$\\
		$s^2(\phi^\dagger \phi)$ & 0 & $\textcolor{blue}{\kappa_s} + \textcolor{blue}{\kappa_{s\phi}}$ & $\textcolor{blue}{\kappa_{s\phi}}$  & $\textcolor{blue}{y_t \kappa_{s\phi}}$ & $\textcolor{blue}{ g_2^2 \kappa_{s\phi} }$ & $0$\\
		$(\phi^\dagger \phi)^2$ & 0 & $0$ & ${\kappa_{s\phi}}$  & $0$ & $0$ & $0$
	\end{tabular}
	\caption{\it Structure of the renormalizable anomalous dimension matrix induced by effective interactions. 
		Terms in blue show the contributions that deviate significantly from naive dimensional analysis; see the text for details.}
	\label{tab:uv4}
\end{table}

The complete expressions for the RGEs are provided in the \texttt{ALPRunner} package~\cite{ALPRunner} published along this work.
Instead, a more graphic picture is presented in Tabs.~\ref{tab:uv5} and~\ref{tab:uv4}, where we show the structure of the anomalous dimension matrices, resulting from one-loop insertions of the SM and new physics (NP) couplings up to $\mathcal{O}\left(1/\Lambda\right)$. 
Each Wilson coefficient in the rows is renormalized by the non-zero coefficients in the columns. 
We have highlighted in blue the contributions which deviate from naive dimensional analysis~\cite{Gavela:2016bzc} by at least one order of magnitude, and which are therefore expected to play an important role in the singlet phenomenology.

Most of the zeros in these tables are trivial. For example, it is not possible to insert an $s^5$ operator in a one-loop diagram without leaving at least 3 external singlet legs, which explains the non-renormalization results in the first column of Tab.~\ref{tab:uv5}. On the other hand, operators with $n$ singlet fields can renormalize others with $m>n$ singlet fields via the insertion of renormalizable NP interactions. Interestingly, both $sXX$ and $sX\widetilde{X}$ operators renormalize the fermionic interactions of the singlet, while $sXX$ renormalizes also the mixed interactions in the scalar sector.

\begin{figure}[t]\centering
	\includegraphics[width=.2\textwidth]{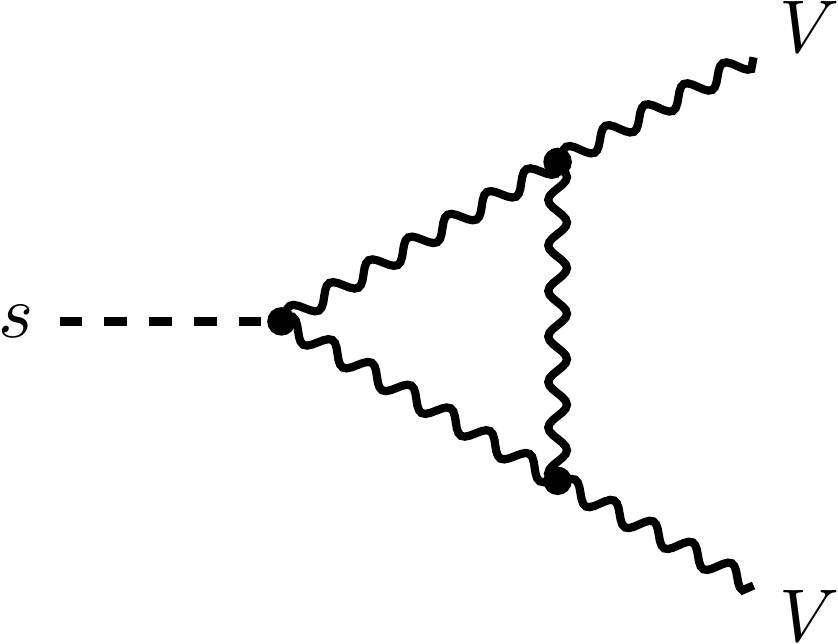}
	\includegraphics[width=.2\textwidth]{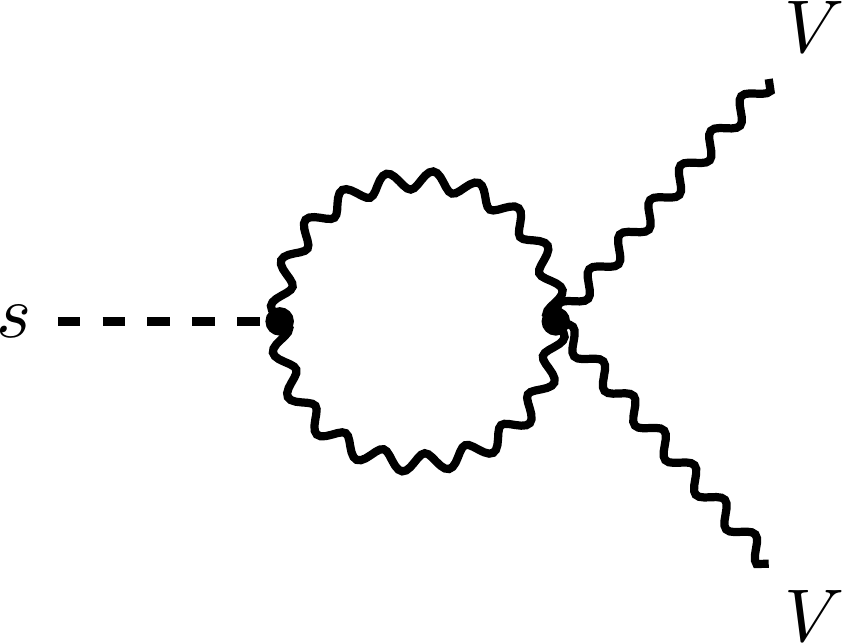}
	\includegraphics[width=.2\textwidth]{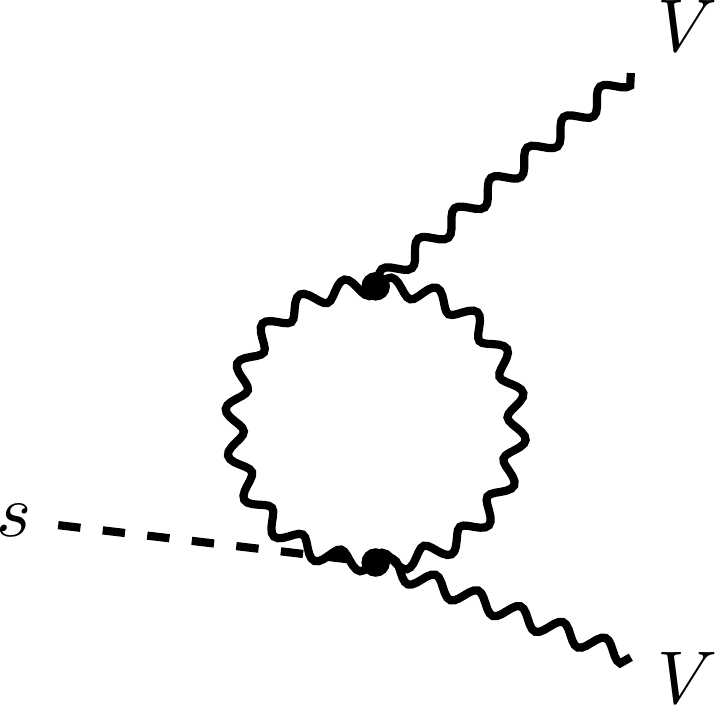}
	\includegraphics[width=.2\textwidth]{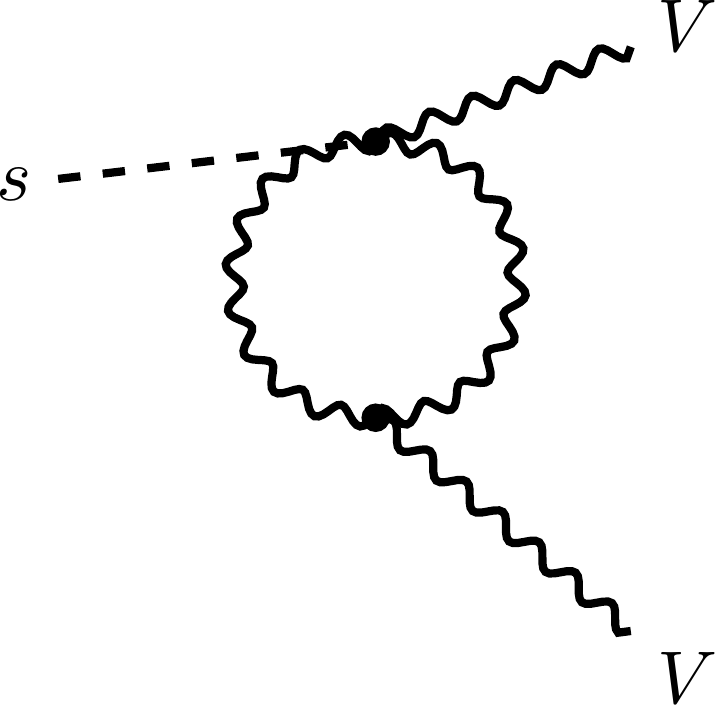}
	\caption{\it One-loop diagrams that contribute to the $s\rightarrow VV$ divergent amplitude, $V$ denoting any SM vector, at order $\mathcal{O}\left( 1/\Lambda \right)$.}
	\label{fig:sWW-diagrams}
	\end{figure}

At one-loop, in the most generic basis, we find that the pseudoscalar interactions with gauge bosons run proportionally to the gauge coupling,
\begin{align}
\beta_{a_{s\widetilde{B},sB}} & =  \frac{41}{3} g_1^2 a_{s\widetilde{B}} (a_{s {B}}) \,,\\
\beta_{a_{s\widetilde{W},sW}} & = - \frac{19}{3} g_2^2 a_{s\widetilde{W}} (a_{s {W}}) \,, \\
\beta_{a_{s\widetilde{G},sG}} & = - 14 g_3^2 a_{s\widetilde{G}} (a_{s {G}}) \,.
\end{align}
%
In fact, if we redefine $a_{sX,\,s\widetilde{X}} \equiv g_X^2 c_{X,\,\widetilde{X}}$, we find that the $c_{X,\,\widetilde{X}}$ couplings do not run at all \cite{Chala:2020wvs,Bauer:2020jbp,Bonilla:2021pvu}. This follows from a non-trivial cancellation of the diagrams shown in Fig.~\ref{fig:sWW-diagrams} and agrees with the considerations presented in the previous section. Indeed, considering the gauge sector in isolation, the CP-even couplings $c_{\widetilde{X}}$ cannot be multiplicatively renormalized since they are multiplied by an effective angle that is $2\pi$ periodic.
The additional pseudoscalar interactions in the EFT could in principle break the scale invariance of these couplings, as they explicitly break this periodicity.
However, no one-loop diagram can be constructed with a single insertion of $s\overline{\psi} \psi \phi$ (or other) terms. 
On the other hand, the CP-odd couplings $c_X$ are not renormalized because $X^2$ is the trace of the conserved energy momentum tensor~\cite{Grojean:2013kd}.

Regarding the singlet-fermion interactions, our results show that they produce at low energies interactions between the pseudoscalar and the Higgs, both at the renormalizable and non-renormalizable levels. These effects are absent in shift-symmetric scenarios, that is for an ALP, as in this case the fermionic interactions grow with momentum and hence can only renormalize themselves~\cite{Bauer:2020jbp}.
Furthermore, the presence of the SM Yukawas in the RGEs of the singlet fermionic couplings implies that CP-even UV setups are not, in general, stable upon running.

\section{Matching at the electroweak scale}\label{sec:matchingbegins}

After the Higgs field develops a VEV, at energies $v\sim 246$\,GeV, the SM particles develop masses. Given the large hierarchy between the masses of the EW bosons and the top quark, in comparison to that of the remaining particles in the SM, the former can be integrated out rendering a simpler low-energy EFT (LEFT). 
While this theory can be fully general and matched to different models above the EW scale, we present in Sec.~\ref{sec:matching} the tree-level matching results assuming that the singlet $s$ is the only BSM $d.o.f.$ in the UV. With this aim, we start by diagonalizing the mass matrix of the scalar sector.

\subsection{Scalar mixing}\label{sec:mixing}
The scalar mass matrix induced by the interactions in our EFT reads 
\begin{equation}
\mathbf{M}_{\rm eff}^2 = \mathbf{M}_0^2 + \mathbf{M}_1^2\,,
\end{equation}
where
\begin{equation}
\mathbf{M}_0^2 =
\begin{pmatrix}
m_s^2 + \frac{\lambda_{s\phi}}{2} v^2 + v_s \left( \kappa_s + \frac{\lambda_s}{2} v_s\right)&  v\left( \kappa_{s\phi} + \lambda_{s\phi} v_s\right) \\
v\left( \kappa_{s\phi} + \lambda_{s\phi} v_s\right)  & -\mu^2 + 3 \lambda v^2 + \kappa_{s\phi} v_s + \frac{\lambda_{s\phi}}{2} v_s^2 
\end{pmatrix}\,,
\label{eq:M2}
\end{equation}
and
\begin{equation}
\mathbf{M}_1^2 =
\begin{pmatrix}
-3 a_{s^3} v^2 v_s - 20 a_{s^5} v_s^3 & -v \left( a_{s} v^2 + 3 a_{s^3} v_s^2\right)\\
-v  \left( a_{s} v^2 + 3 a_{s^3} v_s^2\right) & -v_s  \left(3 a_{s} v^2 + a_{s^3} v_s^2\right)
\end{pmatrix} \,.
\end{equation}
In the expressions above, $v$ is the observed Higgs VEV, determined from muon decay measurements, while $v_s$ denotes the singlet VEV which solves the tadpole equations
\begin{align}
\label{eq:tad1}
\mu^2 & = \lambda v^2  + \frac{1}{2}  v_s \left( 2 \kappa_{s \phi} + \lambda_{s \phi} v_s\right) -  a_{s} v^2 v_s - a_{s^3} v_s^3\,, \\
\label{eq:tad2}
 2 m_s^2 v_s + \kappa_{s\phi} v^2 & =  - \lambda_{s\phi} v^2 v_s  - \kappa_s v_s^2 - \frac{\lambda_s}{3} v_s^3 +\frac{a_s}{2} v^4 + 3 a_{s^3} v^2 v_s^2 + 10 a_{s^5} v_s^4 \,.
\end{align}
Moreover, $(v,\,v_s)$ is guaranteed to be a minimum of the scalar potential if it satisfies the following conditions:
\begin{equation}
\text{Det} (\mathbf{M}_{\rm eff}^2) >0\quad \text{and} \quad \text{Tr} (\mathbf{M}_{\rm eff}^2) >0\,.
\end{equation}

The mass matrix in Eq.~\ref{eq:M2} can be diagonalized by the rotation
\begin{equation}
\begin{pmatrix}
s \\
h
\end{pmatrix} = \begin{pmatrix}
\cos{\theta} & -\sin{\theta}  \\
\sin{\theta}  & \cos{\theta} 
\end{pmatrix} \begin{pmatrix}
\hat{s} \\
\hat{h}
\end{pmatrix}\,,
\end{equation}
with
\begin{equation}
\label{eq:mixingangle}
\tan{2\theta} = \frac{- 2 a_s v^3 +40 a_{s^5} v_s^4/v - \frac{4}{3} v_s/v \left(6 m_s^2 + 3 \kappa_s v_s + \lambda_s v_s^2\right) }{\left(4 a_s - 6a_{s^3}\right) v^2 v_s - 40 a_{s^5} v_s^3 + 2 m_s^2 + \left(-4 \lambda + \lambda_{s\phi}\right) v^2 + v_s\left(2 \kappa_s + \lambda_s v_s\right) 	} \,.
\end{equation}
To obtain this expression, we have used Eqs.~\ref{eq:tad1} and~\ref{eq:tad2} to rewrite $(\mu^2,\,\kappa_{s\phi})$ in terms of the remaining parameters in the EFT. 
In the absence of CP-odd interactions, $v_s = 0$ is a solution of the tadpole equations consistent with the SM assignment of the Higgs parameters, i.e. $\mu^2 = \lambda v^2$. In this limit, the weak and the physical bases of the scalar sector automatically coincide. From the expression above, it is also clear that at the renormalizable level, if the UV parameters are chosen so that $v_s$ vanishes, no mixing with the Higgs boson can arise.  Neglecting the effective interactions, our expressions agree exactly with the results obtained in previous works~\cite{Profumo:2007wc,Beniwal:2018hyi,Ellis:2022lft}. Nonetheless and in spite of the popularity of the singlet framework, there are several studies in the literature exploring the consequences of scalar mixing that neglect the implications of $v_s\neq 0$.

The correlation between the singlet VEV and the scalar mixing is broken by the dimension-five interactions.
Indeed, if the coupling $a_s$ is generated by the UV model, $h-s$ mixing can arise even for $v_s = 0$. This scenario is very particular as it implies an alignment between renormalizable and effective interactions, $\kappa_{s\phi} \sim a_s v^2$, leading to
\begin{equation}
\label{eq:tanVS0}
\tan{2 \theta} \to \frac{2 a_s v^3}{ \left(4 \lambda + \lambda_{s\phi} \right) v^2 - m_s^2}\,.
\end{equation}

Therefore, in full generality, 
we keep track of the $v_s$ dependence in the physical couplings arising from this framework:
\begin{equation}
	\begin{split}
	\label{eq:scalar-after-SSB}
	\mathcal{L}_{s,h}\supset &-\frac{1}{2}\hat{m}_s^2 s^2 - \frac{1}{2}\hat{m}_h^2 h^2 - \frac{\hat{\kappa}_s}{3!}s^3 - \frac{\hat{\kappa}_{s^2 h}}{2}s^2 h - \frac{\hat{\kappa}_{s h^2}}{2}s h^2 - \frac{\hat{\kappa}_h}{3!} h^3 \\ &- \frac{\hat{\lambda}_s}{4!}s^4 - \frac{\hat{\lambda}_{s^3 h}}{3!} s^3 h - \frac{\hat{\lambda}_{s^2 h^2}}{4} s^2 h^2 - \frac{\hat{\lambda}_{s h^3}}{3!} s h^3 - \frac{\hat{\lambda}_h}{4!}h^4 \\
	&+ {\hat{a}_{s^5}}s^5 + {\hat{a}_{s^4 h}}s^4 h + {\hat{a}_{s^3 h^2}}s^3 h^2 + {\hat{a}_{s^2 h^3}}s^2 h^3 + {\hat{a}_{s h^4}}s h^4 + {\hat{a}_{h^5}} h^5\,,
	\end{split}
\end{equation}
where for instance the quartic couplings read as
\begin{align}
\hat{\lambda}_h & =   6\left(\lambda  - a_s v_s + 4 \theta v a_s 	\right)\,,\\
\hat{\lambda}_s & = \lambda_s - 24 \left( \theta v a_{s^3} + 5 v_s a_{s^5}\right) \,,
\end{align}
in the small mixing limit. They are related to the physical scalar masses via the relations:
\begin{align}
\label{eq:mhcap2}
\hat{m}_h^2 & =  2\lambda v^2 -2 a_s v^2 v_s + \theta \frac{2}{v} \bigg[2 m_s^2 v_s + \kappa_s v_s^2 + \frac{\lambda_s}{3} v_s^3 + a_s \frac{v^4}{2} - 10 a_{s^5}v_s^4 \bigg]\,,\\
\hat{m}_s^2 & =   	m_s^2 - \hat{m}_h^2 +2 \lambda v^2 + \frac{\lambda_{s\phi}}{2} v^2  + \frac{\lambda_{s}}{2} v_s^2 + \kappa_s v_s - v_s \bigg[2 a_s v^2 + 3 a_{s^3} v^2 + 20 a_{s}^5 v_s^2\bigg]	\,. 
\end{align}

It is also interesting to point out the relation between the triple and quartic Higgs couplings in the presence of the singlet, taking into account all effective contributions up to $\mathcal{O}(1/\Lambda)$:
\begin{equation}
\hat{\lambda}_h v - \hat{\kappa}_h =\theta \bigg[ \frac{33}{2} v^2 a_s + 30 \frac{v_s^4}{v^2}  a_{s^5}- \frac{v_s}{v^2} \left(6 m_s^2 + 3 \kappa_s v_s + \lambda_s v_s^2\right)\bigg]\,;
\end{equation}
or, in terms of physical parameters,
\begin{equation}
 \frac{1}{2}  \hat{\lambda}_h v - \hat{\kappa}_h =-\frac{3}{2 } \frac{\hat{m}_h^2 }{v}+ 24 \theta  \hat{a}_{s h^4}v^2\,.
\end{equation}
From this expression, it follows that the SM relation between the two physical Higgs couplings can only be broken by non-renormalizable singlet interactions. 

A similar effect occurs in the singlet-Higgs portal couplings. To give an example, in the particular case of $v_s\to 0$, we find that:
\begin{equation}
\hat{\lambda}_{s^2 h^2} - \hat \kappa_{s^2 h}= \theta \bigg[ \kappa_s + v^2 \left( 9 a_{s^3} - 7 a_{s} \right)\bigg]\,,
\end{equation}
where $\theta$ is now determined from Eq.~\ref{eq:tanVS0}. The parameters on the right-hand-side of this equation can be further traded by the physical ones, using the results in App.~\ref{sec:physicalbasisrelations}:
\begin{equation}
\hat{\lambda}_{s^2 h^2} v - \hat \kappa_{s^2 h}= \theta \bigg[\hat \kappa_s + 4 v^2 \left( 6 \hat a_{s^3 h^2} - 7 \hat a_{s h^4} \right)\bigg]\,.
\end{equation}

The gauge sector receives also corrections from a non-vanishing $v_s$. After redefining the gauge couplings,
\begin{equation}
    \label{eq.matching-g1}
    {g}_i \to \hat{g}_i \left(1 -2 v_s a_{s V_i} \right)\,,~\text{with}~\vec{a}_{s V}=\left(a_{sB},\,a_{sW},\,a_{sG}\right) \,,
\end{equation}
as well as the gauge bosons
\begin{equation}
    \label{eq.gauge-redef}
    {V}_i \to \hat{V}_i \left(1 +2 v_s a_{s V_i} \right)\,,
\end{equation}
such that the product $g_i V_i$ remains invariant, 
the physical vector masses read the same as in the SM\,\footnote{Loop effects can however modify these relations; see for example Ref.~\cite{Aiko:2023trb}.}.

Furthermore, the singlet develops renormalizable interactions with the EW gauge bosons due to the mixing with the Higgs particle:
\begin{equation}
	\begin{split}
	\label{eq:gauge-scalar-covariant-derivative-after-SSB}
	\mathcal{L}_{\Phi,V} \supset \sum_{\Phi =s,\,h} & \bigg[\left( c_{\phi Z} +c_{\Phi^2 Z} \Phi\right)\Phi Z^2 + \left(c_{\Phi W} +c_{\Phi^2 W}\Phi\right)\Phi W^2\bigg] \,,
	\end{split}
\end{equation}
where
\begin{align}
    \label{eq.h-covariant-derivative-after-SSB}
    c_{hZ} =& \frac{\hat{g}_2}{2 c_w} m_Z\,, & c_{hW} =& \hat{g}_2 m_W\,, & c_{h^2 Z}=& \frac{\hat{g}_2^2}{8 c_w^2}\,, & c_{h^2 W} =& \frac{\hat{g}_2^2}{4} \,, \\
        c_{sZ} =& \frac{\hat{g}_2}{2 c_w} \theta m_Z\,, & c_{sW} =& \hat{g}_2 \theta m_W\,, & c_{s^2 Z}=& 0\,, & c_{s^2 W} =& 0\,.
\end{align}  
In turn, by mixing with the singlet, the Higgs particle can also couple to~\textit{all} gauge bosons at the non-renormalizable level:
\begin{align}
	\label{eq:gauge-scalar-after-SSB}
	\mathcal{L}_{\Phi,V} \supset 
	\sum_{\Phi =s,\,h}& \bigg[{\hat{a}_{\Phi AA}} \Phi  A_{\mu\nu} A^{\mu\nu} +{\hat{a}_{\Phi ZZ}} \Phi  Z_{\mu\nu} Z^{\mu\nu} + {\hat{a}_{\Phi AZ}} \Phi  A_{\mu\nu} Z^{\mu\nu} + {\hat{a}_{\Phi WW}} \Phi  W_{\mu\nu}^+ W^{-\mu\nu} \nonumber\\ 
	+& {\hat{a}_{\Phi GG}} \Phi  G_{\mu\nu}^A G^{A\mu\nu} 
	+ {\hat{a}_{\Phi A\Tilde{A}}} \Phi  A_{\mu\nu} \widetilde{A}^{\mu\nu} + {\hat{a}_{\Phi Z\Ztil}}\Phi  Z_{\mu\nu} \Ztil^{\mu\nu} + {\hat{a}_{\Phi A\Tilde{Z}}} \Phi  A_{\mu\nu} \widetilde{Z}^{\mu\nu} \bigg.\nonumber\\ 
	\bigg.+& {\hat{a}_{\Phi W\Wtil}}\Phi  W_{\mu\nu}^+ \widetilde{W}^{-\mu\nu} + {\hat{a}_{\Phi G\Gtil}} \Phi  G_{\mu\nu}^A \widetilde{G}^{A\mu\nu} +\dots\bigg]\,, 
	\end{align}
where the ellipsis represent additional 4-body interactions, with at least 2 heavy fields, which will not be important in the discussion that follows. The CP-even couplings above are given by:
\begin{align}
    \label{eq.s-anomalous-couplings-after-SSB}
    \hat{a}_{sAA} =& c_w^2 a_{sB} + s_w^2 a_{sW}  \,, 
    & \hat{a}_{sZZ} =& s_w^2 a_{sB} + c_w^2 a_{sW} \,, \\
    \hat{a}_{sAZ} =& 2 s_w c_w (a_{sW} - a_{sB}) \,, \nonumber
    &  \hat{a}_{sWW} =& 2 a_{sW} \,, \\ \nonumber
     \hat{a}_{sGG} =& a_{sG} \,, \nonumber
     & \hat{a}_{hVV}=& -\theta \hat{a}_{sVV}\,,
\end{align}
while the CP-odd ones are obtained upon the replacement $V\to \widetilde{V}$ and are in agreement with the expressions in the literature~\cite{Bonilla:2022qgm}. 
Note that, in the limit $a_{sW} = a_{sB}$, the singlet coupling to a mixed state of gauge bosons vanishes.
%
%
%

The QCD vacuum also receives contributions from the singlet VEV, as expected from the axion mechanism,
\begin{equation}
    \begin{split}
    \label{eq.theta-term-after-SSB}
        \hat{\theta}_{\rm QCD} =&  \theta_{\rm QCD} + 32\frac{\pi^2}{g_3^2}{a_{s \Gtil}}v_s   \,.
    \end{split}    
\end{equation}

Turning to the fermionic sector, we perform the usual unitary rotations into the mass basis~\cite{LANGACKER1981185}:
\begin{equation}
	\begin{split}
    \label{eq.fermion-rotation}
     \psi_L \rightarrow & V_{\psi_L} \psi_L\,, \\
     \psi_R \rightarrow & V_{\psi_R} \psi_R\,, 
	\end{split}    
\end{equation}
with $\psi = u,d,e$ and $V_{u_L}^\dagger V_{d_L} \equiv V_{\rm CKM}$, $V_{\rm CKM}$ denoting the CKM matrix. The physical Yukawa couplings are then 
\begin{equation}
    \begin{split}
    \label{eq.yukawa-u-after-SSB}
    \hat{y}_{u} =& V_{u_L}^\dagger \left( y_u -i \, {a_{su\phi}v_s}\right)V_{u_R} \,,
    \end{split}    
\end{equation}
\begin{equation}
    \begin{split}
    \label{eq.yukawa-d-after-SSB}
    \hat{y}_{d} =& V_{d_L}^\dagger \left( y_d -i \,  {a_{sd\phi}v_s} \right) V_{d_R} \,,
    \end{split}    
\end{equation}
\begin{equation}
    \begin{split}
    \label{eq.yukawa-e-after-SSB}
    \hat{y}_{e} =& V_{e_L}^\dagger\left( y_e -i \, {a_{se\phi}v_s}\right) V_{e_R}\,,
    \end{split}    
\end{equation}
with the usual relation to the fermion masses:
\begin{equation}
    \begin{split}
    \label{eq.fermion-mass-after-SSB}
    \hat{m}_{\psi} =& \frac{v}{\sqrt{2}} \hat{y}_{\psi} \,.
    \end{split}    
\end{equation}
The corresponding Lagrangian, including the interactions with the scalar sector, reads:
\begin{align}
	\label{eq:scalar-fermion-after-SSB}
	\mathcal{L}_{\Phi,\psi} & \supset 
	\sum_{\psi=u,d,e} \Bigg\{ \overline{\psi} i \,  \slashed{D} \psi - \bigg[ \overline{\psi_L} \hat{m}_{\psi}  \psi_R + h \bar{\psi_L} {\hat{c}_{h \psi}} \psi_R -i \,  s \bar{\psi}_L \hat{c}_{s \psi} \psi_R   \bigg.
 \\ \nonumber
	& -\bigg. 
	s^2 \bar{\psi}_L \hat{a}_{s^2 \psi} \psi_R - {i \, } s h \bar{\psi}_L \hat{a}_{s h \psi} \psi_R- h^2 \bar{\psi}_L \hat{a}_{h^2 \psi} \psi_R + \text{h.c.} \bigg] \Bigg\} \,,
	\end{align}
where
\begin{equation}
	\begin{split}
	\label{eq:s-fermion-after-SSB}
	\hat{c}_{s\psi} =& \frac{1}{\sqrt{2}} V_{\psi_L}^\dagger \Bigg[ i \,  \theta y_\psi + {a_{s\psi\phi}}\left( v + \theta v_s \right)\Bigg] V_{\psi_R} \,,
	\end{split}
\end{equation}
\begin{equation}
	\begin{split}
	\label{eq:h-fermion-after-SSB}
	\hat{c}_{h\psi} =& {\frac{1}{\sqrt{2}} V_{\psi_L}^\dagger \Bigg[ y_\psi -{i \, } {a_{s\psi\phi}} \left( v_s - \theta v\right)\Bigg] V_{\psi_R}}\,,
	\end{split}
\end{equation}
\begin{equation}
	\begin{split}
	\label{eq:s2-fermion-after-SSB}
	\hat{a}_{s^2\psi} =&  V_{\psi_L}^\dagger \frac{i \,  \theta a_{s\psi\phi}}{\sqrt{2}} V_{\psi_R} \,,
	\end{split}
\end{equation}
\begin{equation}
	\begin{split}
	\label{eq:h2-fermion-after-SSB}
	\hat{a}_{h^2\psi} =&  - \hat{a}_{s^2\psi} \,,
	\end{split}
\end{equation}
\begin{equation}
	\begin{split}
	\label{eq:sh-fermion-after-SSB}
	\hat{a}_{s h\psi} =&  V_{\psi_L}^\dagger \frac{{a_{s\psi\phi}}}{\sqrt{2}} V_{\psi_R} \,.
	\end{split}
\end{equation}

\subsection{The low-energy singlet EFT}\label{sec:matching}

%
We can now integrate out the heavy $d.o.f.$ to obtain the LEFT description of the interactions among light particles (including the singlet). To dimension-five, the most general and minimal LEFT Lagrangian can be written as
\begin{align}
	\mathcal{L}_{\text{LEFT}} &= \frac{1}{2} (\partial_{\mu} s) (\partial^{\mu} s) -\frac{1}{2} \Tilde{m}_s^2 s^2 -\frac{\Tilde{\kappa}_{s}}{3!}s^3 - \frac{\Tilde{\lambda}_{s}}{4!} s^4  - \frac{1}{4} G^A_{\mu\nu} G^{A \mu\nu} - \frac{1}{4} A_{\mu\nu} A^{\mu\nu} + \Tilde{\theta}_{\rm QCD} G_{\mu\nu}^A \widetilde{G}^{A\mu\nu} \nonumber\\
	& + \sum\limits_{\psi=u,d,e} \bigg[\overline{\psi} i \,  \slashed{D} \psi -  \overline{\psi_L} \Tilde{m}_\psi \psi_R + i \,  s\overline{\psi_L} \Tilde{c}_\psi \psi_R +s^2 \overline{\psi_L} \Tilde{a}_{\psi} \psi_R + \text{h.c.} \bigg] + {\Tilde{a}_{s^5}}s^5\nonumber\\
		& + {\Tilde{a}_{sA}} s A_{\mu\nu} A^{\mu\nu} + {\Tilde{a}_{sG}}s G_{\mu\nu}^A G^{A\mu\nu} + {\Tilde{a}_{s\Tilde{A}}} s A_{\mu\nu} \widetilde{A}^{\mu\nu} + {\Tilde{a}_{s\Tilde{G}}}s G_{\mu\nu}^A \widetilde{G}^{A\mu\nu} \nonumber\\
		\label{eq:LEFT}
	& + \sum\limits_{\psi=u,d,e} \bigg[  \overline{\psi_L} \Tilde{a}_{\psi A} \sigma^{\mu\nu}\psi_R A_{\mu\nu}+ \overline{\psi_L}\Tilde{a}_{\psi G} \sigma^{\mu\nu} T_A \psi_R G_{\mu\nu}^A +\text{h.c.} \bigg] \,,
\end{align}
where the non-renormalizable couplings are dimensionfull, $\Tilde{a}\equiv \Tilde{a}^0/v$.

The parameters in this tilde basis can be fully fixed at the scale $\mu = v$ by requiring that both theories, before and after EWSB, 
describe the same physics. While the RGEs we will derive next include all the couplings  in Eq.~\ref{eq:LEFT}, below we present the matching equations assuming that the UV model is well described by the SM+$s$ EFT, presented in Sec.~\ref{sec:theory}. At tree level, we obtain:
\begin{equation}
    \begin{split}
    \label{eq.matching-theta-QCD}
    \Tilde{\theta}_{\rm QCD} =& \hat{\theta}_{\rm QCD}\,,
    \end{split}    
\end{equation}
\begin{equation}
    \begin{split}
    \label{eq.matching-kappa-s}
    \Tilde{\kappa}_{s} =\hat{\kappa}_s\,,\quad & \quad \Tilde{m}_{s} = \hat{m}_s\,,
    \end{split}    
\end{equation}
\begin{equation}
    \begin{split}
    \label{eq.matching-lambda-s}
    \Tilde{\lambda}_s=& \hat{\lambda}_s-3\left(\frac{\hat{\kappa}_{s^2 h}}{\hat{m}_h}\right)^2 \,,
    \end{split}    
\end{equation}
\begin{equation}
    \begin{split}
    \label{eq.matching-m}
    \Tilde{m}_{\psi} =\hat{m}_{\psi}\,,\quad & \quad \Tilde{c}_{\psi} = \hat{c}_{s\psi}\,,
    \end{split}    
\end{equation}
\begin{equation}
    \begin{split}
    \label{eq.matching-a-psi}
\Tilde{a}_{\psi}=& \hat{a}_{s^2 \psi} +\frac{\hat{\kappa}_{s^2 h} \hat{c}_{h \psi}}{\hat{m}_h^2}\,,
    \end{split}    
\end{equation}
\begin{equation}
    \begin{split}
    \label{eq.matching-a-s5}
   \Tilde{a}_{s^5}=&\hat{a}_{s^5}+ \frac{7}{5!}\frac{\hat{\kappa}_{s^2h}\hat{\lambda}_{s^3 h}}{\hat{m}_h^2} \,,
    \end{split}    
\end{equation}
\begin{equation}
	\begin{split}
    \label{eq.matching-a-sV}
    \Tilde{a}_{sV}=\hat{a}_{sV}, \quad&\text{and}\quad \Tilde{a}_{s\Tilde{V}}=\hat{a}_{s\widetilde{V}}\,,
	\end{split}    
\end{equation}
with all other Wilson coefficients vanishing. Special care needs to be taken with respect to contributions of $\mathcal{O}(\hat{\kappa}_{s^2 h}/\hat{m}_h^2)$. While naively they seem to be of the same order as that of operators we are neglecting in the LEFT,
in fact ${\hat{\kappa}_{s^2h}/\hat{m}_h^2 \sim \mathcal{O}(1/v)}$ in the small mixing limit. On the contrary, since $\hat{c}_{h \psi} \sim \mathcal{O}(\Tilde{m}_\psi/v)$, the second term in Eq.~\ref{eq.matching-a-psi} should be neglected in our study. Large mixing angles can however spoil this power counting; 
see Eq.~\ref{eq:mhcap2} and App.~\ref{sec:physicalbasisrelations}.

Finally, 
we remark that with the results of the previous section, where we have obtained the full matching to the physical basis including operators with more than one heavy particle, the accuracy of the previous relations could be easily improved beyond tree level.

\section{Low-energy anomalous dimensions}\label{sec:leftadm}

The steps to perform renormalization in the LEFT are the same as described in the HE regime. An important difference is however the set of redundant operators that can be induced, at one loop level, by those in Eq.~\ref{eq:LEFT}; see App.~\ref{sec:greensbases}. We have fixed the one-loop divergences of operators in the Green's basis using~\texttt{matchmakereft} \cite{Carmona:2021xtq}. The subsequent projection onto the minimal basis was performed analytically using the LEFT EOMs~\cite{Chala:2020wvs}. Up to CP-odd terms, the RGEs resulting from this procedure agree exactly with those presented in Ref.~\cite{Chala:2020wvs}, which were obtained with distinct methods.
The full set of equations is written in the \texttt{ALPRunner} package published along this work. Here, we discuss only some contributions that we find particularly interesting.

Contrarily to the results in the HE EFT, in the LEFT the singlet-gauge boson couplings can run; for instance:
\begin{align}
\label{eq:leftsG}
\beta_{\Tilde{a}_{s\widetilde{G}}} & =  \left(-\frac{46}{3} \Tilde{g}_3^2 + 2\text{Tr} \left[\Tilde{c_e} \Tilde{c_e}^\dagger+3\Tilde{c_d} \Tilde{c_d}^\dagger  + 3\Tilde{c_u} \Tilde{c_u}^\dagger    \right] \right)    \Tilde{a}_ {s\widetilde{G}}   \nonumber \\
&~~~~ - 2 g_3 \text{Tr} \left[  \Tilde{a}_{d G} \Tilde{c_d}^\dagger+  \Tilde{a}_{u G} \Tilde{c_u}^\dagger +\text{h.c.} \right]\,,
\end{align}
and similarly for $\beta_{\Tilde{a}_{s G}}$ upon the replacement $\Tilde{a}_{\psi G} \to \ii \Tilde{a}_{\psi G} $.
The first contribution in Eq.~\ref{eq:leftsG} stems entirely from the running of the gauge coupling, such that if all other contributions vanish, $\tilde{c}_{\widetilde{G}} \equiv \Tilde{a}_{s\widetilde{G}}/\tilde{g}_3^2$ remains scale invariant. This occurs if the UV physics above EWSB can be described by the SM+$s$ EFT with $\theta\ll 1$, as in this case $\tilde{c}_\psi$ couplings are $\mathcal{O}(1/\Lambda)$. 
We might still wonder about the additive contributions to this RGE due to the presence of dipole operators, in view of the arguments presented in Sec.~\ref{sec:ALPbasis}, as these operators could be generated by additional field content in the UV without spoiling shift-symmetry. Nonetheless, if this symmetry is assumed, that is in the ALP scenario, $\Tilde{a}_{\psi G} \Tilde{c}_\psi$ terms are at most $\mathcal{O}(\Lambda^{-2})$.

Therefore, up to the EFT expansion order considered in this work, the running of $\Tilde{a}_{s\widetilde{X}}$ couplings is consistent with the expectations discussed in Sec.~\ref{sec:ALPbasis}. To test their robustness at one-loop level,
the inclusion of dimension-six operators in the ALP EFT is required.

Moreover, we find that no contributions to the mass are generated by the LEFT running assuming the singlet is an ALP. 
For a generic pseudoscalar however, apart from multiplicative contributions, the mass can be renormalized by the trilinear coupling $\Tilde{\kappa}_s^2$ as well as by $\tilde{m}_\psi^2 \tilde{c}_\psi^2$ and $\tilde{m}_\psi^3 \tilde{a}_\psi$ terms.

In turn, the singlet mass renormalizes the mass of fermions in the LEFT. This contribution is again absent once matching this theory to the ALP EFT in the UV, up to the expansion order we are considering.

Regarding the running of fermionic couplings, we obtain:
\begin{align}
\label{eq:leftce}
\beta_{\Tilde{c}_{e}} & =  - 6 \Tilde{e}^2 \Tilde{c}_{e} - 24 \Tilde{e}^2 \Tilde{m}_e\Tilde{a}_{s \widetilde{A}}   -24i \Tilde{e}^2  \Tilde{m}_e\Tilde{a}_{s A}  + 2 \text{Tr} \left[	\Tilde{c}_{e} \Tilde{c}_{e}^\dagger + 3\left(\Tilde{c}_{u} \Tilde{c}_{u}^\dagger + \Tilde{c}_{d} \Tilde{c}_{d}^\dagger \right)\right] \Tilde{c}_{e} \\ 
&~~~~ + 3 \Tilde{c}_{e} \Tilde{c}_{e}^\dagger \Tilde{c}_{e} + 2 \text{Tr} \left[\Tilde{m}_e \Tilde{c}_{e}^\dagger \Tilde{a}_{e} - 2 \Tilde{a}_{e} \Tilde{m}_{e}^\dagger \Tilde{c}_{e} + \Tilde{a}_{e} \Tilde{c}_{e}^\dagger \Tilde{m}_{e} - 2 \Tilde{c}_{e} \Tilde{m}_{e}^\dagger \Tilde{a}_{e} \right] \nonumber \\
&~~~~ - 2 i \Tilde{\kappa}_{s} \Tilde{a}_{e} - 12 \Tilde{e}^2 \left(\Tilde{m}_{e} \Tilde{c}_{e}^\dagger \Tilde{a}_{e A} - \Tilde{a}_{e A} \Tilde{m}_{e}^\dagger \Tilde{c}_{e} + \Tilde{a}_{e A} \Tilde{c}_{e}^\dagger \Tilde{m}_{e}	 - \Tilde{c}_{e} \Tilde{m}_{e}^\dagger \Tilde{a}_{e A}		\right)\,, \nonumber
\end{align}
simplifying into the first three terms upon the use of Eqs.~\ref{eq.matching-theta-QCD}--\ref{eq.matching-a-sV} in the absence of scalar mixing. Similarly to this case, $\Tilde{a}_{s V} \Tilde{c}_\psi$ are the only new terms in the RGEs of both $\Tilde{a}_\psi$ and $\Tilde{a}_{\psi V}$ couplings relatively to the CP-even scenario explored in Ref.~\cite{Chala:2020wvs}.

The RGE of the singlet self-coupling is also modified in the presence of CP-odd interactions:
\begin{align}
\label{eq:leftls}
\beta_{\Tilde{\lambda}_{s}} & \supset  -480 \Tilde{\kappa}_s \Tilde{a}_{s^5}+ 60 i \,  \Tilde{\kappa}_s  \text{Tr} \left[ \Tilde{a}_e^\dagger \Tilde{c}_e + 3 \left(\Tilde{a}_u^\dagger \Tilde{c}_u + \Tilde{a}_d^\dagger \Tilde{c}_d \right) +\text{h.c.}\right]\,,
\end{align}
where the large deviations from naive dimensional analysis are apparent. 

Besides, the running of the pure CP-odd couplings in the theory is given by:
\begin{align}
\label{eq:leftks}
\beta_{\Tilde{\kappa}_{s}} & = 3 \Tilde{\kappa}_{s} \Tilde{\lambda}_{s} + 6 \Tilde{\kappa}_{s}\text{Tr} \left[	\Tilde{c}_{e} \Tilde{c}_{e}^\dagger + 3\left(\Tilde{c}_{u} \Tilde{c}_{u}^\dagger + \Tilde{c}_{d} \Tilde{c}_{d}^\dagger \right)\right]   + 24 i \, \text{Tr} \bigg[ \Tilde{m}_e^\dagger  \Tilde{c}_e \Tilde{c}_e^\dagger \Tilde{c}_e + 3 \Tilde{m}_u^\dagger  \Tilde{c}_u \Tilde{c}_u^\dagger \Tilde{c}_u \bigg.\nonumber \\
& \bigg.+ 3\Tilde{m}_d^\dagger  \Tilde{c}_d \Tilde{c}_d^\dagger \Tilde{c}_d + \text{h.c.} \bigg]  -120\Tilde{m}_s^2 \Tilde{a}_{s^5} + 60 i \,  \Tilde{m}_s^2 \text{Tr} \bigg[	\Tilde{c}_{e} \Tilde{a}_{e}^\dagger + 3\left(\Tilde{c}_{u} \Tilde{a}_{u}^\dagger + \Tilde{c}_{d} \Tilde{c}_{a}^\dagger  \right) + \text{h.c.}\bigg] \nonumber\\
& + 24 i \,  \text{Tr}\bigg[\Tilde{m}_{e} \Tilde{m}_{e}^\dagger \Tilde{a}_{e}\Tilde{c}_{e}^\dagger +\Tilde{m}_{e} \Tilde{c}_{e}^\dagger \Tilde{m}_{e} \Tilde{a}_{e}^\dagger + \Tilde{m}_{e} \Tilde{c}_{e}^\dagger \Tilde{a}_{e} \Tilde{m}_{e}^\dagger + 3\left(\Tilde{m}_{u} \Tilde{m}_{u}^\dagger \Tilde{a}_{u}\Tilde{c}_{u}^\dagger +\Tilde{m}_{u} \Tilde{c}_{u}^\dagger \Tilde{m}_{u} \Tilde{a}_{u}^\dagger \right)\bigg.\nonumber\\
& \bigg. + 3\left( \Tilde{m}_{u} \Tilde{c}_{u}^\dagger \Tilde{a}_{u} \Tilde{m}_{u}^\dagger + \Tilde{m}_{d} \Tilde{m}_{d}^\dagger \Tilde{a}_{d}\Tilde{c}_{d}^\dagger +\Tilde{m}_{d} \Tilde{c}_{d}^\dagger \Tilde{m}_{d} \Tilde{a}_{d}^\dagger + \Tilde{m}_{d} \Tilde{c}_{d}^\dagger \Tilde{a}_{d} \Tilde{m}_{d}^\dagger\right) + \text{h.c.}  \bigg]\,;
\end{align}
\begin{align}
\label{eq:leftas5}
\beta_{\Tilde{a}_{s^5}} & = 10 \Tilde{a}_{s^5}  \text{Tr} \left[	\Tilde{c}_{e} \Tilde{c}_{e}^\dagger + 3\left(\Tilde{c}_{u} \Tilde{c}_{u}^\dagger + \Tilde{c}_{d} \Tilde{c}_{d}^\dagger \right)\right] + \frac{i}{3}  \, \Tilde{\lambda}_s \text{Tr} \left[	\Tilde{a}_{e} \Tilde{c}_{e}^\dagger + 3\left(\Tilde{a}_{u} \Tilde{c}_{u}^\dagger + \Tilde{a}_{d} \Tilde{c}_{d}^\dagger \right) + \text{h.c.}\right]  \nonumber \\
&+10 \Tilde{\lambda}_s \Tilde{a}_{s^5} + 4 i \,  \text{Tr} \left[\Tilde{a}_{e}^\dagger \Tilde{c}_{e} \Tilde{c}_{e}^\dagger \Tilde{c}_{e} + 3\left(\Tilde{a}_{u}^\dagger \Tilde{c}_{u} \Tilde{c}_{u}^\dagger \Tilde{c}_{u} +\Tilde{a}_{d}^\dagger \Tilde{c}_{d} \Tilde{c}_{d}^\dagger \Tilde{c}_{d}  \right) + \text{h.c.}  \right]\,.
\end{align}

\section{\texttt{ALPRunner} usage: phenomenological application}\label{sec:application}

In this section, we show the non-trivial impact of running in phenomenological studies of the singlet by computing the strongest EDM constraints on particular UV scenarios (where only two BSM couplings are assumed non-zero). 
We use the EDM expressions obtained in Ref.~\cite{PhysRevD.104.095027}
\,\footnote{We found a mismatch by a factor of 4 in the expression for the fermionic contribution to the $G\widetilde{G}G$ Weinberg operator, with respect to the results in Ref.~\cite{Dekens:2014jka}, the latter leading to stronger bounds.}. 
To include all the running effects computed in this work, namely contributions from scalar mixing, we take into account the energy evolution of the couplings involved in such expressions\,\footnote{To avoid double counting, the $\log^2$ contributions obtained in~\cite{PhysRevD.104.095027}, that arise from the (partial) renormalization of the singlet couplings, are not considered.}.

Before presenting the results, we illustrate below the main features of the new \texttt{ALPRunner} package.
Our EDM analyses are based on dedicated routines which can be generalized to scan the HE parameter space and impose bounds on more general UV scenarios than those considered in this section. The functions available are enumerated in Tab.~\ref{tab:ALPRunner0}. 

Users can download the package from the link in Ref.~\cite{ALPRunner} 
and initialize it in the following way:

\vspace{0.5cm}
\begin{mmaCell}{Input}
Needs["ALPRunner'"]
\end{mmaCell}
\begin{mmaCell}{Output}

\end{mmaCell}
{%
\begin{figure}[h]
\vspace{-3em}
\hspace{2.5cm} \includegraphics[scale=.1]{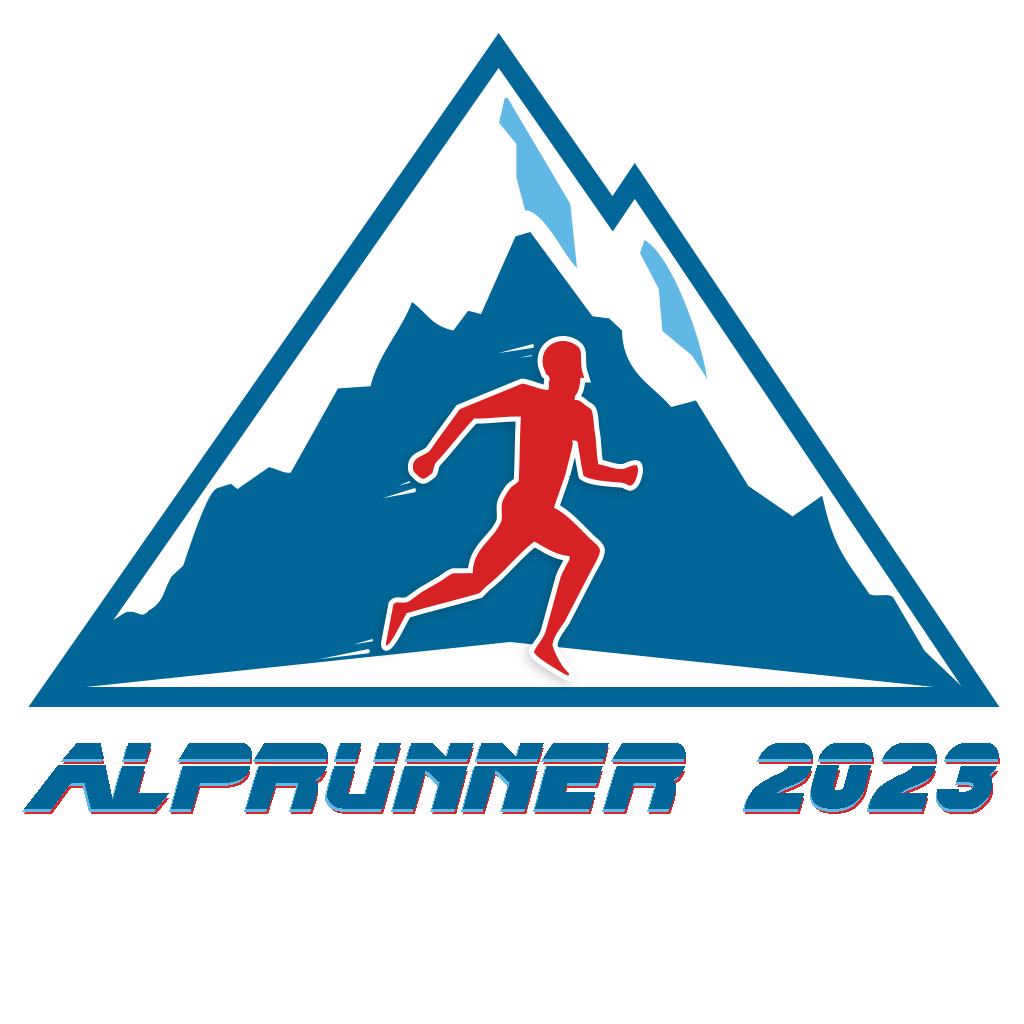}
\vspace{-2em}
\end{figure}

\hspace{2cm} \texttt{SDB, JMR, MR.}

\hspace{2cm} \texttt{arXiv:2306.08036}\\

}%

The complete RGE of a given Wilson coefficient can be obtained via the function `\texttt{rge\$variable\_name}' (e.g.~\texttt{rge\$asB}). 
Additional functions have been included to solve numerically the running equations for user-given boundary conditions (BCs). In the code line below, the SM BCs (defined at the scale $\Lambda^\prime = 172$ GeV) are stored in the list `\texttt{smLOWBC}'. The following function finds then the values of the SM couplings at a user-set UV scale ($\Lambda =10^3$\,GeV in the example below), which are compatible with the given low-scale BCs. 
%
\begin{mmaCell}[moredefined={smLOWBC}]{Input}
\mmaDef{rgeSMto\(\Lambda\)reset}[smLOWBC,\mmaSup{10}{3},172];
\end{mmaCell}
%
%

The BSM BCs are stored in a different list, `\texttt{bsmBC}', defined at the $\Lambda$ scale. The nomenclature used for the Wilson coefficients in the package is the same as in Eq.~\ref{eq:basis}. Moreover, we have included a function `\texttt{basisChange}' to translate operators from the derivative basis in Eq.~\ref{eq:ALPL} to the EFT basis used in this work. After defining the BCs, the function `\texttt{rgePSEFTparamsolvereset}' solves the HE RGEs parametrically\,\footnote{If the user wishes to fix numerically the BCs for \textit{all} BSM parameters, the function `\texttt{rgePSEFTreset[bsmBC,1000,172]}' should be called instead.
}. For instance, in the example below, two UV parameters remain free, $a_{s \widetilde{B}}\equiv$ \texttt{asBt} and ${a_{s B}\equiv \texttt{asB}}$:

\begin{mmaCell}[moredefined={bsmBC,rgePSEFTparamsolvereset,bsmHSparameters, asBt, asB}]{Input}
rgePSEFTparamsolvereset[bsmBC,\mmaSup{10}{3},172,
                        \{\{asBt,asBtparam\},\{asB,asBparam\}\}];
  
\end{mmaCell}

In the next step, we show how to match automatically the HE and the LE EFT couplings, as described in Sec.~\ref{sec:matchingbegins}.
 In this procedure, the UV values of the Higgs coupling and scale, $\mu$ and $\lambda$, are fixed in order to comply with the experimental constraints\,\footnote{This module is discussed in detail in the notebook file `\texttt{ALPRunner\_Guide.nb}'~\cite{ALPRunner}.} $v=246\,$GeV and $\hat{m}_h (v) = 125\,$GeV~\cite{Workman:2022ynf}. The singlet VEV is then directly obtained from Eqs.~\ref{eq:tad1} and~\ref{eq:tad2}, as well as the mixing angle defined in Eq.~\ref{eq:mixingangle}. Points $(v,\,v_s)$ which are not minima of the scalar potential are discarded.

Under these conditions and with fixed $v_s$, the LE BCs are set by the function `\texttt{bcLE}', as written below.
\begin{mmaCell}[moredefined={boundaryConditionLE,PSLEFTmatching, bcLE, rgePSEFTparamsolve}]{Input}
boundaryConditionLE = bcLE[PSLEFTmatching[172],
			rgePSEFTparamsolve,\{2*\mmaSup{10}{-6},3*\mmaSup{10}{-6}\}];
		
\end{mmaCell}
The first argument in this function requires the matching conditions (defined internally), which are defined in terms of the UV couplings that result from solving parametrically the HE RGEs, given in the second argument. The latter have been evaluated for specific values of $(a_{s \widetilde{B}},\,a_{s B}) = (2\times 10^{-6},\,3\times 10^{-6})$.

Given the LE BCs, the following function solves numerically the LE RGEs, outputing the EFT parameters at a given IR scale (chosen below to be 5\,GeV):
\begin{mmaCell}[moredefined={rgeLEreset, boundaryConditionLE}]{Input}
rgeLEreset[boundaryConditionLE,172,5];
\end{mmaCell}
After these steps, any LE Wilson coefficients can be determined at a scale $<v$, e.g.
\begin{mmaCell}[moredefined={WCOsAt,rgeLE}]{Input}
WCOsAt[5.2]/.rgeLE
\end{mmaCell}
\begin{mmaCell}{Output}
1.47055*\mmaSup{10}{-6}
\end{mmaCell}

\begin{figure}[t!]
	\centering
	\includegraphics[scale=0.28]{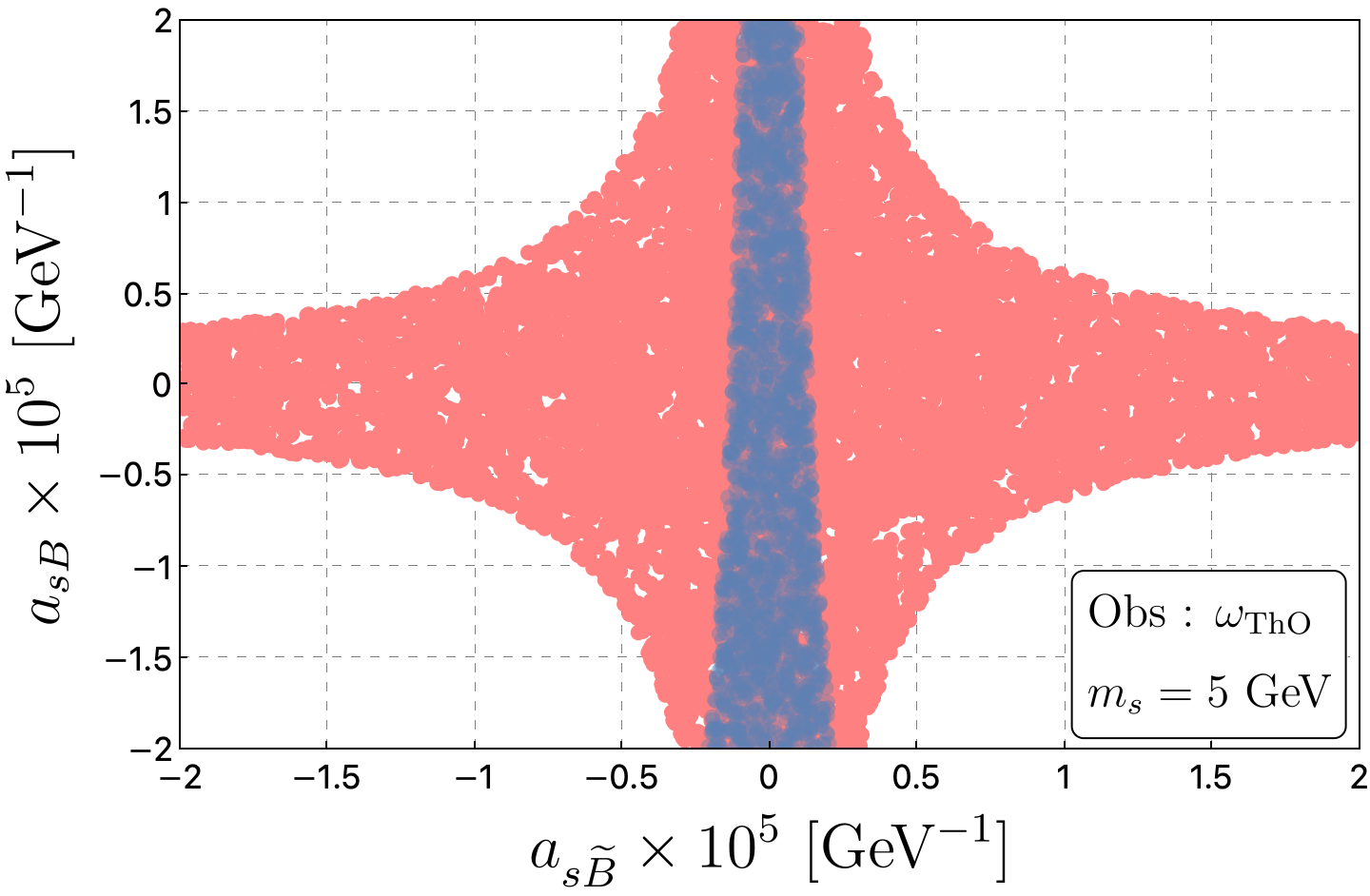} \quad
	\includegraphics[scale=0.28]{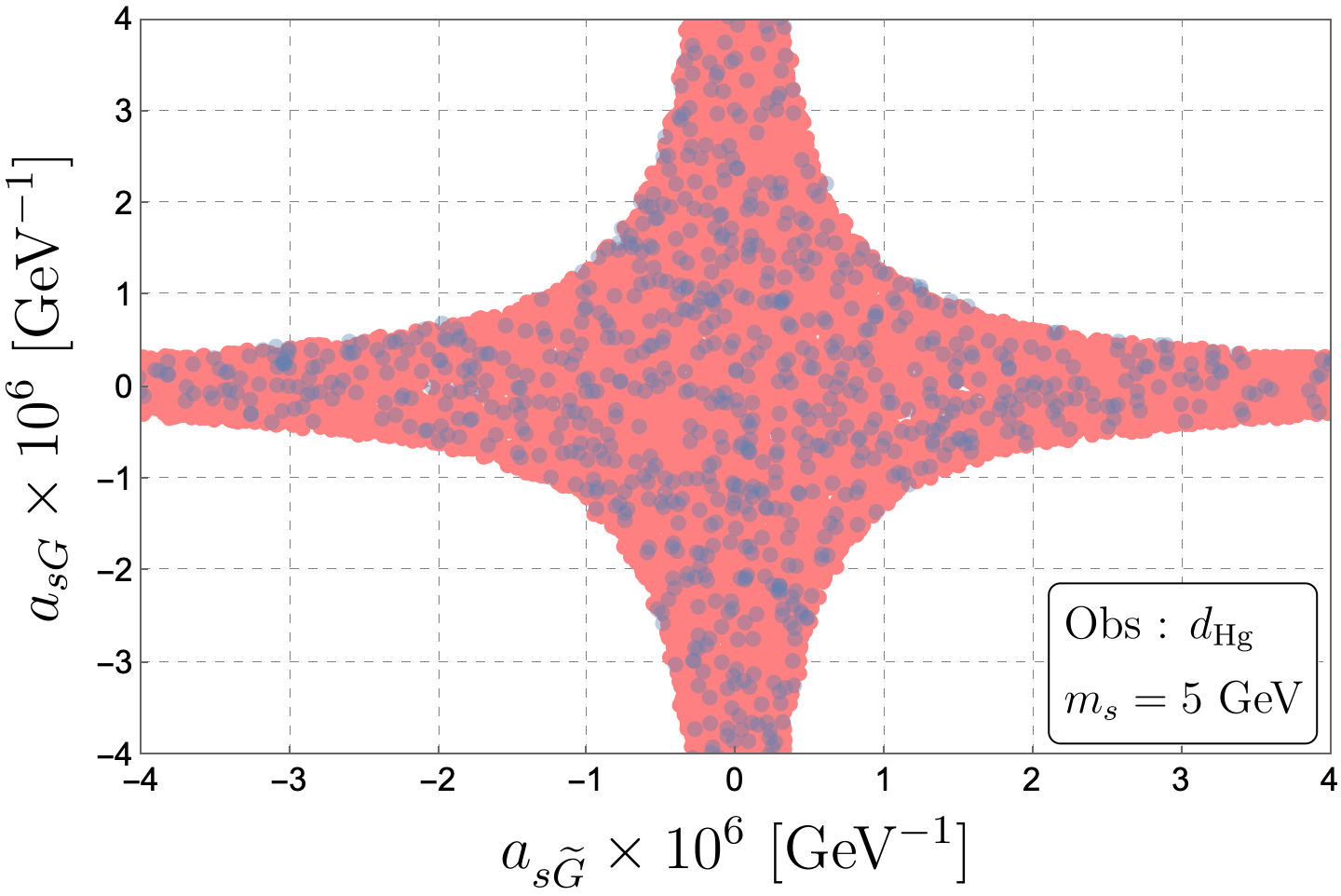}
	\caption{\it EDM constraints on the UV gauge scenarios discussed in the text for $\Lambda = 1$\, TeV and $m_s = 5$\,GeV. The red points are allowed in the scenarios with vanishing $\kappa_{s\phi}$, but are further constrained to the blue region for $\kappa_{s\phi} = -0.1$.  The running of all couplings in the EFT is considered, from the scale $\Lambda$ down to $\mu=5\,$GeV. The leading observable setting the bounds is identified in the plot legend.}
	\label{fig:EDM-sVV}
\end{figure}

\begin{figure}[t]
	\centering
	\includegraphics[scale=0.4]{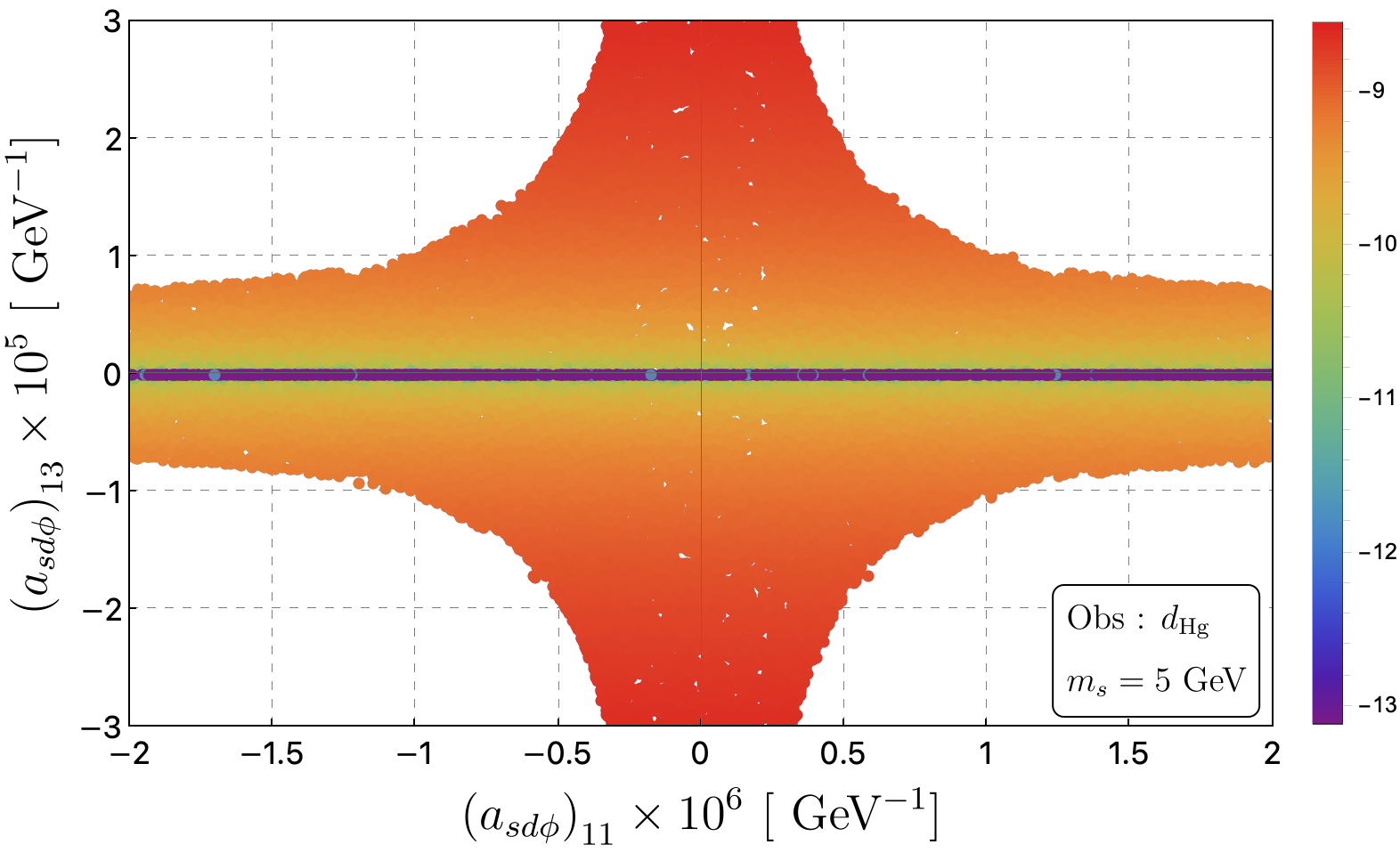}
	\caption{\it EDM constraints on the fermionic UV scenario discussed in the text for ${\Lambda = 1\,\text{TeV}}$ and $m_s = 5$\,GeV. The leading observable setting the bounds is identified in the plot legend. The colour bar identifies the size (in log scale) of the shift-breaking invariants; see Eq.~\ref{eq:INV}.}\label{fig:EDM-sFF}
\end{figure}

If the user is interested in generating several UV points, to understand namely which of those are bounded by IR constraints, the following scan function can be used:
\begin{mmaCell}[moredefined={scanHEspace,bsmBC,bsmHSparameters,asBt, asB, Observable},index=100]{Input}
scanHEspace[bsmBC,1000,172,5,
            \{\{asBt,asBtparam\},\{asB,asBparam\}\},
            \{10000,\{-5*\mmaSup{10}{-6},+5*\mmaSup{10}{-6}\},\{-5*\mmaSup{10}{-6},+5*\mmaSup{10}{-6}\}\},
            Observable, "File.m"];
            
\end{mmaCell}
where the first argument is the list of HE BCs; the second, third and fourth arguments are, respectively, the UV, matching and IR scales; the fifth is the list of UV parameters that the user chooses to let free (and which are not defined in \texttt{bsmBC}); 
whereas the sixth entry corresponds to the number and interval of random points to be generated. Finally, the two last arguments encode the expression --- defined by the user --- that will be evaluated at the IR scale, for each point that was generated, and the file where the information will be stored. In the analyses presented below, `\texttt{Observable}' is a list of expressions of the different contributions to EDMs induced by the singlet interactions~
--- defined in Eq. 9 of Ref.~\cite{PhysRevD.104.095027} --- that we have subsequently compared with the relevant experimental bounds, in order to constrain a given UV scenario. Some results are shown in Figs.~\ref{fig:EDM-sVV} and~\ref{fig:EDM-sFF}.

\vspace{0.25cm}
\textbf{UV gauge scenarios.} Assuming that only $(a_{s\widetilde{B}},\,a_{sB})$ are generated by the UV, the constraints presented in the left panel of  Fig.~\ref{fig:EDM-sVV} imply that
\begin{equation}
\label{eq:boundBB}
|a_{s\widetilde{B}}\,a_{sB}| \lesssim~6 \times 10^{-11} \text{ GeV}^{-2}\,,
\end{equation}
for $m_s = 5\,$GeV.
This bound is set by measurements of the electron EDM using the molecule ThO~\cite{THO} and agrees with the results reported in Ref.~\cite{PhysRevD.104.095027}. The results are largely insensitive to $m_s$. 

Tree level mixing effects can have a non-negligible contribution to the fermion coupling and therefore modify the bound obtained above. For instance, assuming $\kappa_{s\phi} \sim -0.1$, which leads to 
$\theta\sim \mathcal{O}(10^{-3})$, we find a stronger constraint on the gauge couplings as represented by the blue points in Fig.~\ref{fig:EDM-sVV}, which turns out to be largely independent of $a_{s B}$. This is because, for a (red) point in the positive quadrant that was allowed by the previous analysis, the $\theta$ contribution in Eq.~\ref{eq:s-fermion-after-SSB} dominates over the others induced by running. As such, the EDM contribution, which essentially goes as ${\rm Re}[\tilde c_e]  a_{sB} + {\rm Im}[\tilde c_e ] a_{s\widetilde B}$ becomes a function of $a_{s \widetilde B}$ only. However, the width of the new allowed region is not exactly constant and more points are allowed in the $a_{sB}<0$ area. This is due to a possible cancellation between the imaginary terms in Eq.~\ref{eq:s-fermion-after-SSB}, that is, those induced by running vs. those induced by scalar mixing.

In the right panel of Fig.~\ref{fig:EDM-sVV}, we have explored another UV scenario where only the singlet-gluon couplings are assumed non-zero. The leading observable setting the bounds in this case is the neutron EDM of the $^{199}$Hg atom~\cite{PhysRevLett.116.161601}, which requires
 \begin{equation}
\label{eq:boundGG}
|a_{s\widetilde{G}}\,a_{sG}| \lesssim~1.3 \times 10^{-12} \text{ GeV}^{-2}\,,
\end{equation}
for $m_s = 5\,$GeV. This bound is within the same order of magnitude than the one obtained in Ref.~\cite{PhysRevD.104.095027}\,\footnote{We however do not consider the running of the EDM Lagrangian, Eq. (9) in Ref.~\cite{PhysRevD.104.095027}, from 5 GeV down to 1 GeV.} and is insensitive to the value of $\kappa_{s\phi}$. 

\vspace{0.25cm}
\textbf{UV fermionic scenarios.} In Fig.~\ref{fig:EDM-sFF}, we instead explored a UV fermionic scenario where the only non-vanishing couplings are $(a_{sd\phi})_{11}$ and $(a_{sd\phi})_{13}$, assumed to take real values\,\footnote{We work in a UV basis where $V_{d_L} = V_{\rm CKM}$ and $V_{u_L} = 1$.}. This setup aims to illustrate an important point: 
strong EDM bounds can be applied to some CP-even BSM scenarios. The CP violation induced in this case stems from the mixing of the CP-even physics with the SM Yukawa couplings\,\footnote{A general study on the generation of this ``opportunistic'' CP violation was presented in Ref.~\cite{Bonnefoy:2023bzx}, in the framework of the SM EFT.}, induced by the fermion mass diagonalization and the RGEs (see Tab.~\ref{tab:uv5}).
As in the previous case, the leading observable setting the bounds is the neutron EDM. 

In Fig.~\ref{fig:EDM-sFF}, we have also represented the size of the shift-breaking invariants\,\footnote{These invariants are defined in Eq.~(5)~of~Ref.~\cite{Bonnefoy:2022rik}.}, i.e.
\begin{equation}
\label{eq:INV}
I_{\Sigma} \equiv \frac{1}{\Lambda} \sqrt{(I_{d}^1)^2 + (I_{d}^2)^2 + (I_{d}^3)^2}\,,
\end{equation}
for each point represented in the plot. 
We see that the most shift-symmetric points lie close to the horizontal axis. In particular, ALPs 
lie exactly in the blind direction of the plot, since the UV texture 
\begin{equation}
a_{sd\phi} (\Lambda) = \begin{pmatrix}
\times & 0 & \times \\
0& 0 & 0 \\
0 & 0 & 0
\end{pmatrix}
\end{equation}
is not compatible with an exact shift-symmetry. Using~\texttt{ALPRunner}, it is straightforward to generalize this flavour matrix and obtain meaningful bounds in the ALP parameter space.

\section{Conclusions}\label{sec:conclusion}

In this work, we have computed the energy evolution of all couplings in the most general EFT obtained by extending the SM with a real (pseudo) scalar singlet. Our results are valid up to one-loop accuracy and order $1/\Lambda$ in the EFT expansion. The renormalization group equations have been furthermore included in a new Mathematica package -- \texttt{ALPRunner} -- with functions to solve the running numerically, 
as well as to match directly the high-energy with the low-energy EFT couplings, assuming that the singlet is the only BSM $d.o.f.$ at energies above the EW scale.

We have found that CP-odd couplings in the singlet EFT are, in general, renormalized by CP-even ones due to the mixing with the Yukawa couplings of the SM. Namely, CP-even fermiophilic scenarios can generate, at low energies, CP-odd couplings between the singlet and the Higgs boson, both at the renormalizable and effective levels. 
This is possible due to the RGE mixing of operators of different mass dimensions, which is a common feature of both the high-energy and low-energy EFTs. Moreover, 
BSM operators can renormalize pure SM ones, such as the Higgs quartic coupling. 
At the matching level, we have computed all effective contributions to scalar mixing, which modify important relations among the physical couplings. In particular, we found that via this mixing and only at dimension-five, the SM relation between the triplet and quartic Higgs couplings can be distorted. Furthermore, it is possible for the singlet to have a vanishing VEV and still mix with the Higgs boson, an effect that is absent in the renormalizable framework, and which could be of phenomenological relevance. (While mixing effects are expected to be small from experimental searches~\cite{SinBound}, the current interpretations assume correlations that do not necessarily hold at the effective level. A re-interpretation of the experimental analyses is therefore necessary to quantify the actual size of these effects\,\footnote{For example, new contributions to the coupling $s^2 h$ (see Eq.~\ref{eq:sh2-after-SSB}) could potentially lead to new regions of the parameter space compatible with larger values of $\theta$.}.) 

Throughout this work, we have also discussed the shift-symmetric limit of our results and showed that they are in perfect agreement with the quantization rules imposed by the ALP periodicity. Conversely, for a generic singlet, the couplings to gauge bosons can in general run and the interpretation of the experimental bounds must take this running effect into account.

Finally, as an application of our results, we obtained EDM bounds on particular UV scenarios. We have shown that the constraints on the singlet couplings to the abelian gauge boson can be significantly impacted by the presence of scalar mixing. 
Generic constraints on the singlet parameter space can be obtained for more general UV setups using the package presented along this work, as we have explained in Sec.~\ref{sec:application}. The diagonalization of the scalar and fermion systems has been implemented in the package, as well as the full dependence of the EFT coefficients on the scalar VEVs. An algorithm was included to make sure the UV inputs comply with EW constraints. 
Finally, the shift-breaking invariants~\cite{Bonnefoy:2022rik} of the singlet EFT have also been defined in the package, such that the phenomenology of a more shift-symmetric singlet, or that of an ALP, can be studied in full detail with \texttt{ALPRunner}.

\vspace{1cm}

\section*{Acknowlegments}
We are grateful to Quentin Bonnefoy, Mikael Chala, V\'ictor Enguita, Renato Fonseca, Bel\'en Gavela, Jonathan Kley, Javier Fuentes-Mart\'in, Luca Merlo  and Jorge Fern\'andez de Troc\'oniz for illuminating discussions, and Jos\'e Santiago and Pablo Olgoso for helping with \texttt{matchmakereft} \cite{Carmona:2021xtq}. We thank Luca Di Luzio and Paride Paradisi for pointing out a misunderstanding in the computation of the EDM contributions.
We thank Anisha and Guilherme Guedes for comments on the manuscript, and Sunando Patra for suggestions on Mathematica utilities, and functions imported (and modified) from \texttt{CoDEx} \cite{DasBakshi:2018vni}. The work of M. R. is supported by the Marie Sklodowska-Curie grant agreement No 860881-HIDDeN. The work of J.M.R. is supported by the Spanish MICIU through the National Program FPI-Severo Ochoa (grant number PRE2019-089233). S.D.B. acknowledges financial support from the Spanish Research Agency (Agencia Estatal de Investigación) under the grants  No. PID2019-106087GB-C21/ 10.13039/501100011033 and PID2021-128396NB-100/AEI/10.13039/501100011-
\noindent 033; by the Junta de Andalucía (Spain) under the grants No \ FQM- 101, A-FQM-467-UGR18 and P18-FR-4314 (FEDER).  J.M.R. acknowledges partial financial support from the Spanish Research Agency through the grants IFT Centro de excelencia Severo Ochoa Program No CEX2020-001007-S and PID2019-108892RB-I00, funded by MCIN/AEI/ 10.13039/50110001103. The logo of \texttt{ALPRunner} was inspired by output from DALL.E 2 of OpenAI.

\newpage
\appendix

\section{\texttt{ALPRunner} toolbox}\label{app:toolbox}

\vspace{1cm}
\begin{table}[h!]
	\centering
	\begin{threeparttable}
		\bf
		\footnotesize
		\noindent\begin{tcolorbox}[bgtable=logo-final.png,tabularx*={}{|cl},watermark stretch=1,width=1.05\textwidth,watermark opacity=.1,colback=white!35,opacityback=.95]
			\hline \hline
			\rule{0pt}{1.5\normalbaselineskip} \textbf{Function} &  \textbf{Details}  \\[2mm]
			\hline \hline
			\rule{0pt}{1.5\normalbaselineskip} \texttt{\color{blue} rg\$variablename} &  Prints the RGE of \texttt{variablename}. 
			\\[2mm]
			& 
			\texttt{Names["rg\$*"]} returns the list of all RGE variables.\\[2mm]
			\hline
			\rule{0pt}{1.5\normalbaselineskip} \texttt{\color{blue}rgePSEFTreset}  &  Solves numerically the RGEs in the unbroken-phase. \\[2mm]
			& The output is saved in \texttt{rgePSEFT}.\\[2mm]
			\hline		
			\rule{0pt}{1.5\normalbaselineskip} \texttt{\color{blue}rgePSEFTparamsolvereset}  &  Solves parametrically the RGEs in the unbroken-phase. \\[2mm]
			& The output is saved in \texttt{rgePSEFTparamsolve}. \\[2mm]
			\hline
			\rule{0pt}{1.5\normalbaselineskip} \texttt{\color{blue}PSLEFTmatching}  &  Matches the unbroken and broken phase EFT couplings.\\[2mm]
			\hline
			\rule{0pt}{1.5\normalbaselineskip} \texttt{\color{blue}bcLE}  &  Creates the boundary conditions for the \texttt{rgeLEreset} function.\\[2mm]
			\hline
			\rule{0pt}{1.5\normalbaselineskip} \texttt{\color{blue}rgeLEreset}  &  Solves numerically the RGEs in the broken-phase. \\[2mm]
			&  The output is saved in \texttt{rgeLE}.\\[2mm]
			\hline
			\rule{0pt}{1.5\normalbaselineskip} \texttt{\color{blue}scanHEspace}  &  Scans a multi-dimensional parameter space and accepts/ \\[2mm]
			& rejects the points given (user-defined) observables.\\[2mm]
			\hline
			\rule{0pt}{1.5\normalbaselineskip} \texttt{\color{blue}basisChange}  &  Translates couplings defined in the derivate basis to the\\[2mm]
			&  one used in this work.\\[2mm]
			\hline
			\rule{0pt}{1.5\normalbaselineskip} \texttt{\color{blue}Iu1}  & Evaluates the shift invariant structure `$I_u^1$'. All 14 invariants \\[2mm]
			& are defined, and are called by `Iu1', `Id1', `Ie1', `Iud1', etc.\\[2mm]
			\hline
			\hline
		\end{tcolorbox}
	\end{threeparttable}
\caption{Main functions provided by \texttt{ALPRunner}. For more details, see `\texttt{ALPRunner\_Guide.nb}' in~\cite{ALPRunner}.}
\label{tab:ALPRunner0}
\end{table}
\newpage

\section{Green's bases}\label{sec:greensbases}
\begin{table}[h]
	\centering
\renewcommand{\arraystretch}{1.4}
\scalebox{0.95}{
\begin{tabular}{|c|c|c|c|}
		\hline \hline
		Scalar & Yukawa & Gauge & Derivative \\ \hline
       $\mathcal{O}_{s^5} = s^5$ & $\mathcal{O}_{su\phi}= i s \overline{q_L}\Tilde{\phi}u_R $ & $\mathcal{O}_{s\widetilde{B}}=s B_{\mu\nu} \widetilde{B}^{\mu\nu}$ & $\mathcal{R}_{s\phi\Box} = is \phi^\dagger D^2 \phi$ \\
        $\mathcal{O}_{s^3} = s^3 (\phi^\dagger \phi)$ & $\mathcal{O}_{sd\phi}= i s \overline{q_L}\phi d_R $ &  $\mathcal{O}_{s\widetilde{W}}=s W_{\mu\nu} \widetilde{W}^{\mu\nu}$ & $\mathcal{R}_{s\Box} = s^2 \partial^2 s$ \\
       $\mathcal{O}_{s} = s (\phi^\dagger \phi)^2 $ & $\mathcal{O}_{se\phi}= i s \overline{l_L}\phi e_R $ &  $\mathcal{O}_{s\widetilde{G}}=s G_{\mu\nu} \widetilde{G}^{\mu\nu}$ & $\mathcal{R}_{\phi s\Box} = \phi^\dagger \phi \partial^2 s$ \\
        & & $\mathcal{O}_{s B}=s B_{\mu\nu} B^{\mu\nu}$ & $\mathcal{R}_{s q} = i s \overline{q_L} \slashed{D}  q_L$ \\
        & & $\mathcal{O}_{s W}=s W_{\mu\nu} W^{\mu\nu}$ & $\mathcal{R}_{s l} = i s \overline{l_L} \slashed{D}  l_L$ \\
        & & $\mathcal{O}_{s G}=s G_{\mu\nu} G^{\mu\nu}$ & $\mathcal{R}_{s u} = i s \overline{u_R} \slashed{D}  u_R$ \\
        & & & $\mathcal{R}_{s d} = i s \overline{d_R} \slashed{D}  d_R$ \\
        & & & $\mathcal{R}_{s e} = i s \overline{e_R} \slashed{D}  e_R$ \\
    \hline \hline
\end{tabular}
}
\caption{Green's basis of the SM+$s$ EFT at dimension-five~\cite{Chala:2020wvs}. 
The operators labelled with $\mathcal{R}$ are redundant when evaluating an amplitude on-shell.}\label{tab:greensbasis}
\end{table}
\qquad

\begin{table}[htb!]
	\centering
\renewcommand{\arraystretch}{1.4}
\scalebox{0.95}{
\begin{tabular}{|c|c|c|c|c|}
		\hline \hline
		Scalar & Yukawa & Gauge & Dipole & Derivative \\ \hline
		$\tilde{\mathcal{O}}_{s^5} = s^5$ & $\tilde{\mathcal{O}}_{u}=  s^2 \overline{u_L}u_R$ & $\tilde{\mathcal{O}}_{s\widetilde{A}}=s A_{\mu\nu} \widetilde{A}^{\mu\nu}$  & $\tilde{\mathcal{O}}_{uA}=\overline{u_L} \sigma^{\mu\nu}u_R A_{\mu\nu} $ & $\tilde{\mathcal{R}}_{s\Box} = s^2 \partial^2 s$  \\
        & $\tilde{\mathcal{O}}_{d}=  s^2 \overline{d_L}d_R$ &  $\tilde{\mathcal{O}}_{s\widetilde{G}}=s G_{\mu\nu} \widetilde{G}^{\mu\nu}$  & $\tilde{\mathcal{O}}_{dA}=\overline{d_L} \sigma^{\mu\nu}d_R A_{\mu\nu} $  & $\tilde{\mathcal{R}}_{u \Box} = \overline{u_L} D^2 u_R$ \\
        & $\tilde{\mathcal{O}}_{e}=  s^2 \overline{e_L}e_R$ &  $\tilde{\mathcal{O}}_{s A}=s A_{\mu\nu} A^{\mu\nu}$  & $\tilde{\mathcal{O}}_{eA}=\overline{e_L} \sigma^{\mu\nu}e_R A_{\mu\nu} $  & $\tilde{\mathcal{R}}_{d \Box} = \overline{d_L} D^2 d_R$  \\
        & & $\tilde{\mathcal{O}}_{s G}=s G_{\mu\nu} G^{\mu\nu}$  & $\tilde{\mathcal{O}}_{u G}=\overline{u_L} \sigma^{\mu\nu} T_A u_R G_{\mu\nu}^A$ & $\tilde{\mathcal{R}}_{e \Box} = \overline{e_L} D^2 e_R$  \\
        & &   & $\tilde{\mathcal{O}}_{d G}=\overline{d_L} \sigma^{\mu\nu} T_A d_R G_{\mu\nu}^A$ & $\tilde{\mathcal{R}}_{s u_L} = i s \overline{u_L} \slashed{D}  u_L$ \\
        & & & $\tilde{\mathcal{O}}_{e G}=\overline{e_L} \sigma^{\mu\nu} T_A e_R G_{\mu\nu}^A$ & $\tilde{\mathcal{R}}_{s d_L} = i s \overline{d_L} \slashed{D}  d_L$  \\
        & & & & $\tilde{\mathcal{R}}_{s e_L} = i s \overline{e_L} \slashed{D}  e_L$ \\
        & & & & $\tilde{\mathcal{R}}_{s u_R} = i s \overline{u_R} \slashed{D}  u_R$ \\
        & & & & $\tilde{\mathcal{R}}_{s d_R} = i s \overline{d_R} \slashed{D}  d_R$ \\
        & & & & $\tilde{\mathcal{R}}_{s e_R} = i s \overline{e_R} \slashed{D}  e_R$ \\
    \hline \hline
\end{tabular}
}
\caption{Green's basis of the SM+$s$ LEFT at dimension-five~\cite{Chala:2020wvs}. 
The operators labeled with $\mathcal{R}$ are redundant when evaluating an amplitude on-shell.}\label{tab:greensbasisLEFT}
\end{table}

\newpage    
\section{Scalar couplings in the physical basis}\label{sec:physicalbasisrelations}

The scalar interactions defined in the physical basis of Eq.~\ref{eq:scalar-after-SSB} read as follows, in the small mixing angle limit:
\begin{equation}
	\begin{split}
	\label{eq:s3-after-SSB}
	\hat{\kappa}_{s} =& \kappa_s + v_s \lambda_s + 3\theta v \lambda_{s\phi} -3 \Big[ v \left( v + 6 \theta v_s \right) a_{s^3} + 20 v_s^2 a_{s^5}\Big]\,,
	\end{split}
\end{equation}
\begin{equation}
	\begin{split}
	\label{eq:s2h-after-SSB}
	\hat{\kappa}_{s^2 h} =& -\theta \kappa_s +2\theta \kappa_{s\phi} -\theta v_s \lambda_{s} + \left( v +2\theta v_s \right) \lambda_{s\phi} \\\
	+& 3 \Bigg\{ \Big[ \theta \left( v^2 -2v_s^2 \right) -2 v_s v \Big] a_{s^3} -2\theta v^2 a_s + 20 \theta v_s^2 a_{s^5} \Bigg\} \,,
	\end{split}
\end{equation}
\begin{equation}
	\begin{split}
	\label{eq:sh2-after-SSB}
	\hat{\kappa}_{sh^2} =& \kappa_{s\phi} + 6\theta v \lambda + \left(v_s -2\theta v \right)\lambda_{s\phi} -3 \Big[ v \left( 2\theta v_s +v \right)a_s + v_s \left( v_s -4\theta v \right)a_{s^3}\big] \,,
	\end{split}
\end{equation}
\begin{equation}
	\begin{split}
	\label{eq:h3-after-SSB}
	\hat{\kappa}_{h} =& 3\Bigg\{ 2 v \lambda -\theta \kappa_{s\phi} -\theta v_s \lambda_{s\phi} + \Big[ v \left( 3\theta v -2 v_s \right)a_s +3\theta v_s^2 a_{s^3} \Big] \Bigg\} \,,
	\end{split}
\end{equation}
\begin{equation}
	\begin{split}
	\label{eq:s4-after-SSB}
	\hat{\lambda}_{s} =& \lambda_s - 24 \left( \theta v a_{s^3} + 5 v_s a_{s^5}\right) \,,
	\end{split}
\end{equation}
\begin{equation}
	\begin{split}
	\label{eq:s3h-after-SSB}
	\hat{\lambda}_{s^3 h} =& -\theta \lambda_s + 3\theta \lambda_{s\phi} + 6 \Big[ -\left(3\theta v_s + v \right)a_{s^3} + 20 \theta v_s a_{s^5} \Big]  \,,
	\end{split}
\end{equation}
\begin{equation}
	\begin{split}
	\label{eq:s2h2-after-SSB}
	\hat{\lambda}_{s^2 h^2} =& \lambda_{s\phi} - 6 \Big[ 2\theta v a_s + \left( v_s - 2\theta v \right)a_{s^3} \Big] \,,
	\end{split}
\end{equation}
\begin{equation}
	\begin{split}
	\label{eq:sh3-after-SSB}
	\hat{\lambda}_{s h^3} =& 3\Bigg\{ 2\theta \lambda - \theta \lambda_{s\phi} + 2\Big[ - \left( \theta v_s + v \right) a_s + 3\theta v_s a_{s^3} \Big] \Bigg\} \,,
	\end{split}
\end{equation}
\begin{equation}
	\begin{split}
	\label{eq:h4-after-SSB}
	\hat{\lambda}_{h} =& 6\Big[ \lambda - a_s \left( v_s -4 \theta v \right)\Big] \,,
	\end{split}
\end{equation}
\begin{equation}
	\begin{split}
	\label{eq:s5-after-SSB}
	\hat{a}_{s^5} =& a_{s^5}  \,,
	\end{split}
\end{equation}
\begin{equation}
	\begin{split}
	\label{eq:s4h-after-SSB}
	\hat{a}_{s^4 h} =& \theta \left( a_{s^3} -5 a_{s^5} \right)  \,,
	\end{split}
\end{equation}
\begin{equation}
	\begin{split}
	\label{eq:s3h2-after-SSB}
	\hat{a}_{s^3 h^2} =& \frac{a_{s^3}}{2} \,,
	\end{split}
\end{equation}
\begin{equation}
	\begin{split}
	\label{eq:s2h3-after-SSB}
	\hat{a}_{s^2 h^3} =& \theta \left( a_s - \frac{3}{2} a_{s^3} \right)  \,,
	\end{split}
\end{equation}
\begin{equation}
	\begin{split}
	\label{eq:sh4-after-SSB}
	\hat{a}_{s h^4} =& \frac{a_{s}}{4} \,,
	\end{split}
\end{equation}
\begin{equation}
	\begin{split}
	\label{eq:h5-after-SSB}
	\hat{a}_{h^5} =& - \frac{\theta}{4} a_s \,.
	\end{split}
\end{equation}

\bibliographystyle{unsrt}
\bibliography{Bibliography.bib}

\begin{thebibliography}{10}

\bibitem{Bauer:2017ris}
Martin Bauer, Matthias Neubert, and Andrea Thamm.
\newblock {Collider Probes of Axion-Like Particles}.
\newblock {\em JHEP}, 12:044, 2017.

\bibitem{Kaplan:1983fs}
David~B. Kaplan and Howard Georgi.
\newblock {SU(2) x U(1) Breaking by Vacuum Misalignment}.
\newblock {\em Phys. Lett. B}, 136:183--186, 1984.

\bibitem{Kaplan:1983sm}
David~B. Kaplan, Howard Georgi, and Savas Dimopoulos.
\newblock {Composite Higgs Scalars}.
\newblock {\em Phys. Lett. B}, 136:187--190, 1984.

\bibitem{Gripaios:2009pe}
Ben Gripaios, Alex Pomarol, Francesco Riva, and Javi Serra.
\newblock {Beyond the Minimal Composite Higgs Model}.
\newblock {\em JHEP}, 04:070, 2009.

\bibitem{Chala:2012af}
Mikael Chala.
\newblock {$h \rightarrow \gamma\gamma$ excess and Dark Matter from Composite
  Higgs Models}.
\newblock {\em JHEP}, 01:122, 2013.

\bibitem{Vecchi:2013bja}
Luca Vecchi.
\newblock {The Natural Composite Higgs}.
\newblock 2013.

\bibitem{Bellazzini:2014yua}
Brando Bellazzini, Csaba Csáki, and Javi Serra.
\newblock {Composite Higgses}.
\newblock {\em Eur. Phys. J. C}, 74(5):2766, 2014.

\bibitem{Cacciapaglia:2019bqz}
Giacomo Cacciapaglia, Gabriele Ferretti, Thomas Flacke, and Hugo Serôdio.
\newblock {Light scalars in composite Higgs models}.
\newblock {\em Front. in Phys.}, 7:22, 2019.

\bibitem{McDonald:1993ex}
John McDonald.
\newblock {Gauge singlet scalars as cold dark matter}.
\newblock {\em Phys. Rev. D}, 50:3637--3649, 1994.

\bibitem{Burgess:2000yq}
C.~P. Burgess, Maxim Pospelov, and Tonnis ter Veldhuis.
\newblock {The Minimal model of nonbaryonic dark matter: A Singlet scalar}.
\newblock {\em Nucl. Phys. B}, 619:709--728, 2001.

\bibitem{Arcadi:2019lka}
Giorgio Arcadi, Abdelhak Djouadi, and Martti Raidal.
\newblock {Dark Matter through the Higgs portal}.
\newblock {\em Phys. Rept.}, 842:1--180, 2020.

\bibitem{Profumo:2007wc}
Stefano Profumo, Michael~J. Ramsey-Musolf, and Gabe Shaughnessy.
\newblock {Singlet Higgs phenomenology and the electroweak phase transition}.
\newblock {\em JHEP}, 08:010, 2007.

\bibitem{Espinosa:2011eu}
Jose~R. Espinosa, Ben Gripaios, Thomas Konstandin, and Francesco Riva.
\newblock {Electroweak Baryogenesis in Non-minimal Composite Higgs Models}.
\newblock {\em JCAP}, 01:012, 2012.

\bibitem{Chala:2016ykx}
Mikael Chala, Germano Nardini, and Ivan Sobolev.
\newblock {Unified explanation for dark matter and electroweak baryogenesis
  with direct detection and gravitational wave signatures}.
\newblock {\em Phys. Rev. D}, 94(5):055006, 2016.

\bibitem{Ellis:2022lft}
John Ellis, Marek Lewicki, Marco Merchand, Jos\'e~Miguel No, and Mateusz Zych.
\newblock {The scalar singlet extension of the Standard Model: gravitational
  waves versus baryogenesis}.
\newblock {\em JHEP}, 01:093, 2023.

\bibitem{WITTEN1984351}
Edward Witten.
\newblock Some properties of $\mathrm{O}$(32) superstrings.
\newblock {\em Physics Letters B}, 149(4):351--356, 1984.

\bibitem{BANKS1996173}
Tom Banks and Michael Dine.
\newblock Couplings and scales in strongly coupled heterotic string theory.
\newblock {\em Nuclear Physics B}, 479(1):173--196, 1996.

\bibitem{Peccei:1977ur}
R.~D. Peccei and Helen~R. Quinn.
\newblock {Constraints Imposed by CP Conservation in the Presence of
  Instantons}.
\newblock {\em Phys. Rev. D}, 16:1791--1797, 1977.

\bibitem{Peccei:1977hh}
R.~D. Peccei and Helen~R. Quinn.
\newblock {CP Conservation in the Presence of Instantons}.
\newblock {\em Phys. Rev. Lett.}, 38:1440--1443, 1977.

\bibitem{Weinberg:1977ma}
Steven Weinberg.
\newblock {A New Light Boson?}
\newblock {\em Phys. Rev. Lett.}, 40:223--226, 1978.

\bibitem{Wilczek:1977pj}
Frank Wilczek.
\newblock {Problem of Strong $P$ and $T$ Invariance in the Presence of
  Instantons}.
\newblock {\em Phys. Rev. Lett.}, 40:279--282, 1978.

\bibitem{Chala:2020wvs}
Mikael Chala, Guilherme Guedes, Maria Ramos, and Jose Santiago.
\newblock {Running in the ALPs}.
\newblock {\em Eur. Phys. J. C}, 81(2):181, 2021.

\bibitem{Bauer:2020jbp}
Martin Bauer, Matthias Neubert, Sophie Renner, Marvin Schnubel, and Andrea
  Thamm.
\newblock {The Low-Energy Effective Theory of Axions and ALPs}.
\newblock {\em JHEP}, 04:063, 2021.

\bibitem{Bonilla:2021pvu}
J.~Bonilla, I.~Brivio, M.~B. Gavela, and V.~Sanz.
\newblock {One-loop corrections to ALP couplings}.
\newblock {\em JHEP}, 11:168, 2021.

\bibitem{Bonnefoy:2022rik}
Quentin Bonnefoy, Christophe Grojean, and Jonathan Kley.
\newblock {Shift-Invariant Orders of an Axionlike Particle}.
\newblock {\em Phys. Rev. Lett.}, 130(11):111803, 2023.

\bibitem{ALPRunner}
Supratim Das~Bakshi, Jonathan Machado, and Maria Ramos.
\newblock \url{https://github.com/sdbakshi13/ALPRunner}.

\bibitem{OConnell:2006rsp}
Donal O'Connell, Michael~J. Ramsey-Musolf, and Mark~B. Wise.
\newblock {Minimal Extension of the Standard Model Scalar Sector}.
\newblock {\em Phys. Rev. D}, 75:037701, 2007.

\bibitem{Barger:2007im}
Vernon Barger, Paul Langacker, Mathew McCaskey, Michael~J. Ramsey-Musolf, and
  Gabe Shaughnessy.
\newblock {LHC Phenomenology of an Extended Standard Model with a Real Scalar
  Singlet}.
\newblock {\em Phys. Rev. D}, 77:035005, 2008.

\bibitem{Gripaios:2016xuo}
Ben Gripaios and Dave Sutherland.
\newblock {An operator basis for the Standard Model with an added scalar
  singlet}.
\newblock {\em JHEP}, 08:103, 2016.

\bibitem{GEORGI198673}
Howard Georgi, David {B. Kaplan}, and Lisa Randall.
\newblock Manifesting the invisible axion at low energies.
\newblock {\em Physics Letters B}, 169(1):73--78, 1986.

\bibitem{Agrawal:2022yvu}
Prateek Agrawal, Kim~V. Berghaus, JiJi Fan, Anson Hook, Gustavo
  Marques-Tavares, and Tom Rudelius.
\newblock {Some open questions in axion theory}.
\newblock In {\em {Snowmass 2021}}, 3 2022.

\bibitem{Fraser:2019ojt}
Katherine Fraser and Matthew Reece.
\newblock {Axion Periodicity and Coupling Quantization in the Presence of
  Mixing}.
\newblock {\em JHEP}, 05:066, 2020.

\bibitem{Brivio:2017ije}
I.~Brivio, M.~B. Gavela, L.~Merlo, K.~Mimasu, J.~M. No, R.~del Rey, and
  V.~Sanz.
\newblock {ALPs Effective Field Theory and Collider Signatures}.
\newblock {\em Eur. Phys. J. C}, 77(8):572, 2017.

\bibitem{Alloul:2013bka}
Adam Alloul, Neil~D. Christensen, C\'eline Degrande, Claude Duhr, and Benjamin
  Fuks.
\newblock {FeynRules 2.0 - A complete toolbox for tree-level phenomenology}.
\newblock {\em Comput. Phys. Commun.}, 185:2250--2300, 2014.

\bibitem{Hahn:2000kx}
Thomas Hahn.
\newblock {Generating Feynman diagrams and amplitudes with FeynArts 3}.
\newblock {\em Comput. Phys. Commun.}, 140:418--431, 2001.

\bibitem{Hahn:2016ebn}
Thomas Hahn, Sebastian Pa\ss{}ehr, and Christian Schappacher.
\newblock {FormCalc 9 and Extensions}.
\newblock {\em PoS}, LL2016:068, 2016.

\bibitem{Carmona:2021xtq}
Adrian Carmona, Achilleas Lazopoulos, Pablo Olgoso, and Jose Santiago.
\newblock {Matchmakereft: automated tree-level and one-loop matching}.
\newblock {\em SciPost Phys.}, 12(6):198, 2022.

\bibitem{Abbott:1981ke}
L.~F. Abbott.
\newblock {Introduction to the Background Field Method}.
\newblock {\em Acta Phys. Polon. B}, 13:33, 1982.

\bibitem{Gavela:2016bzc}
B.~M. Gavela, E.~E. Jenkins, A.~V. Manohar, and L.~Merlo.
\newblock {Analysis of General Power Counting Rules in Effective Field Theory}.
\newblock {\em Eur. Phys. J. C}, 76(9):485, 2016.

\bibitem{Grojean:2013kd}
Christophe Grojean, Elizabeth~E. Jenkins, Aneesh~V. Manohar, and Michael Trott.
\newblock Renormalization group scaling of higgs operators and h $\rightarrow$
  $\gamma$$\gamma$ decay.
\newblock {\em JHEP}, 04:016, 2013.

\bibitem{Beniwal:2018hyi}
Ankit Beniwal, Marek Lewicki, Martin White, and Anthony~G. Williams.
\newblock {Gravitational waves and electroweak baryogenesis in a global study
  of the extended scalar singlet model}.
\newblock {\em JHEP}, 02:183, 2019.

\bibitem{Aiko:2023trb}
Masashi Aiko and Motoi Endo.
\newblock {Electroweak precision test of axion-like particles}.
\newblock {\em JHEP}, 05:147, 2023.

\bibitem{Bonilla:2022qgm}
J.~Bonilla, A.~de~Giorgi, B.~Gavela, L.~Merlo, and M.~Ramos.
\newblock {The cost of an ALP solution to the neutral B-anomalies}.
\newblock {\em JHEP}, 02:138, 2023.

\bibitem{LANGACKER1981185}
Paul Langacker.
\newblock Grand unified theories and proton decay.
\newblock {\em Physics Reports}, 72(4):185--385, 1981.

\bibitem{PhysRevD.104.095027}
Luca Di~Luzio, Ramona Gr\"ober, and Paride Paradisi.
\newblock Hunting for $\mathrm{CP}$-violating axionlike particle interactions.
\newblock {\em Phys. Rev. D}, 104:095027, Nov 2021.

\bibitem{Dekens:2014jka}
W.~Dekens, J.~de~Vries, J.~Bsaisou, W.~Bernreuther, C.~Hanhart, Ulf-G.
  Mei\ss{}ner, A.~Nogga, and A.~Wirzba.
\newblock {Unraveling models of CP violation through electric dipole moments of
  light nuclei}.
\newblock {\em JHEP}, 07:069, 2014.

\bibitem{Workman:2022ynf}
R.~L. Workman and Others.
\newblock {Review of Particle Physics}.
\newblock {\em PTEP}, 2022:083C01, 2022.

\bibitem{THO}
V.~Andreev and N.~Hutzler.
\newblock Improved limit on the electric dipole moment of the electron.
\newblock {\em Nature}, 562, 10 2018.

\bibitem{PhysRevLett.116.161601}
B.~Graner, Y.~Chen, E.~G. Lindahl, and B.~R. Heckel.
\newblock Reduced limit on the permanent electric dipole moment of
  $^{199}\mathrm{Hg}$.
\newblock {\em Phys. Rev. Lett.}, 116:161601, Apr 2016.

\bibitem{Bonnefoy:2023bzx}
Quentin Bonnefoy, Emanuele Gendy, Christophe Grojean, and Joshua~T. Ruderman.
\newblock {Opportunistic CP Violation}.
\newblock 2 2023.

\bibitem{SinBound}
Richard~Keith Ellis et~al.
\newblock {Physics Briefing Book}: {Input for the European Strategy for
  Particle Physics Update 2020}.
\newblock 10 2019.

\bibitem{DasBakshi:2018vni}
Supratim Das~Bakshi, Joydeep Chakrabortty, and Sunando~Kumar Patra.
\newblock {CoDEx: Wilson coefficient calculator connecting SMEFT to UV theory}.
\newblock {\em Eur. Phys. J. C}, 79(1):21, 2019.

\end{thebibliography}

\end{document}